\documentclass{sig-alternate-mod}
\usepackage{mathptmx}
\usepackage{fancyhdr}
\usepackage[normalem]{ulem}
\usepackage[hyphens]{url}
\usepackage[sort,nocompress]{cite}
\usepackage[final]{microtype}
\usepackage[keeplastbox]{flushend}
\usepackage{tabularx}
\usepackage{threeparttable}
\usepackage{enumitem}
\usepackage{upgreek}
\usepackage{makecell}
\usepackage{multirow}
\usepackage{makecell}
\usepackage{textgreek}
\usepackage[bookmarks=true,breaklinks=true,letterpaper=true,colorlinks,linkcolor=black,citecolor=blue,urlcolor=black]{hyperref}

\newcommand{\insitu}{\textit{in situ}}
\newcommand{\Insitu}{\textit{In situ}}
\newcommand{\etal}{\textit{et al.}}
\newcommand{\mat}[1]{\mathbf{#1}}

\fancypagestyle{firstpage}{
  \fancyhf{}
  
  \fancyfoot[C]{\thepage}
}

\title{On the accuracy of analog neural network inference accelerators} 

\numberofauthors{1}
\author{
\alignauthor
T. Patrick Xiao$^\dagger$, Ben Feinberg$^\dagger$, Christopher H. Bennett$^\dagger$, Venkatraman Prabhakar$^*$, Prashant Saxena$^*$, Vineet Agrawal$^*$, Sapan Agarwal$^\dagger$, Matthew J. Marinella$^{\dagger,\ddagger}$\\[10pt]
\affaddr{$^\dagger$Sandia National Laboratories,  $^*$Infineon Technologies, $^\ddagger$Arizona State University}\\
$^\dagger$\affaddr{\{txiao, bfeinbe, cbennet, sagarwa\}@sandia.gov} \\
$^*$\affaddr{\{Venkatraman.Prabhakar, Prashant.Saxena, Vineet.Agrawal\}@infineon.com}\\
$^\ddagger$\affaddr{m@asu.edu}
}

\begin{document}
\maketitle
\thispagestyle{firstpage}
\pagestyle{plain}

\begin{abstract}
Specialized accelerators have recently garnered attention as a method to reduce the power consumption of neural network inference. A promising category of accelerators utilizes non-volatile memory arrays to both store weights and perform \insitu{} analog computation inside the array.
While prior work has explored the design space of analog accelerators to optimize performance and energy efficiency, there is seldom a rigorous evaluation of the accuracy of these accelerators. This work shows how architectural design decisions, particularly in mapping neural network parameters to analog memory cells, influence inference accuracy.
When evaluated using ResNet50 on ImageNet, the resilience of the system to analog non-idealities---cell programming errors, analog-to-digital converter resolution, and array parasitic resistances---all improve when analog quantities in the hardware are made \emph{proportional} to the weights in the network.
Moreover, contrary to the assumptions of prior work, nearly equivalent resilience to cell imprecision can be achieved by fully storing weights as analog quantities, rather than spreading weight bits across multiple devices, often referred to as bit slicing.
By exploiting proportionality, analog system designers have the freedom to match the precision of the hardware to the needs of the algorithm, rather than attempting to guarantee the same level of precision in the intermediate results as an equivalent digital accelerator. This ultimately results in an analog accelerator that is more accurate, more robust to analog errors, and more energy-efficient.
\end{abstract}

\section{Introduction}
\label{sec:intro}
Deep neural networks (DNNs) have grown rapidly in importance in the past decade, enabling image recognition, natural language processing, predictive analytics, and many other tasks to be performed with high accuracy and generalizability~\cite{Lecun15}. As the size and complexity of DNNs have grown to tackle more challenging problems, so has the demand for increasingly powerful and energy-efficient processors. Hardware that is optimized for DNN processing, which is dominated by matrix operations\cite{Coates13}, has been a major enabler of machine learning innovation. But new, more efficient hardware approaches are needed to keep pace with the rapid developments in artificial intelligence and its growing computational needs~\cite{Xu18}.

Accelerators based on \insitu{} computing---utilizing memory for both storage and computation---have attracted significant attention as a possible path to order-of-magnitude improvements in energy efficiency~\cite{Bojnordi16,Shafiee16,Chi16}. 
These systems harness the analog properties of non-volatile memory arrays to perform many concurrent multiply-and-accumulate (MAC) operations, enabling the computation of a matrix-vector multiplication (MVM) in a single step.

While analog processing offers intrinsic efficiency benefits, it has historically struggled with accuracy. Unlike digital systems, the solution quality in analog systems is directly degraded by noise, process variations, and various parasitic effects. To provide precision on par with digital systems, many prior analog inference accelerators adopt a hybrid approach known as \emph{bit slicing}, where weight values are spread bitwise across multiple memory devices, and the analog intermediate results are digitized and aggregated~\cite{Bojnordi16,Shafiee16,Chi16}. This technique allows weights to be represented more precisely even with low-precision memory devices, but at a higher energy cost than a purely analog approach. Recent work has optimized the performance and energy of bit-sliced accelerators~\cite{Nag18,Ankit19,Chou19,Li20}, but rarely evaluates the effect of system-level design decisions on inference accuracy.

This work studies how architecture affects accuracy in analog inference accelerators. We use a detailed accuracy model for \insitu{} MVMs that includes the effect of various analog errors at the resolution of individual MACs, such as memory cell process variations and array parasitic resistances. The model allows an architectural design space exploration that uses the error sensitivity of end-to-end inference accuracy as the primary figure-of-merit. To provide a sensitivity analysis that can be applied to realistic applications, accuracy is evaluated on ImageNet classification with the ResNet50 neural network from the MLPerf Inference benchmark. This model is also used to benchmark digital systems~\cite{Reddi20}.

Though the accuracy of analog accelerators has been studied~\cite{Sze19}, the analysis in this work provides a more comprehensive view of how accuracy fits into analog architecture design. This work demonstrates that bit slicing offers a smaller benefit than often assumed and typically does not justify its energy cost; moreover, contrary to the assumptions of prior work, bit slicing cannot be used as a mitigation for highly error-prone analog devices. Just as important, when signed arithmetic is handled in analog, it is possible to obtain a linear or \emph{proportional mapping} between the numerical values in the algorithm and the physical quantities that represent them in the analog hardware. This proportionality is the key to enable high inference accuracy and greater robustness to analog errors. Following the end-to-end principle of Saltzer \etal{}~\cite{Saltzer84}, this robustness allows hardware requirements (such as the array size and the analog-to-digital converter resolution) to be relaxed while still ensuring high end-to-end accuracy of an application neural network. For state-of-the-art inference applications, proportionality produces significant improvements simultaneously in accuracy, error tolerance, energy efficiency, and area.

This paper is organized as follows. Section \ref{sec:background} introduces analog inference accelerators, and surveys the architectural design space established by prior work in the field. Section \ref{sec:key_ideas} lays out what we view to be key principles for designing analog systems that can achieve high accuracy and resilience to errors. These conclusions are supported by the results in the remainder of the paper, which are based on the methodology in Section \ref{sec:methods}. Sections \ref{sec:errors}, \ref{sec:ADC}, \ref{sec:quant} and \ref{sec:parasitics} discuss how device- and architecture-level design decisions influence an accelerator's sensitivity to memory cell programming errors, ADC quantization errors, and parasitic resistance errors. Section \ref{sec:sonos} presents a case study of these principles using an exemplar analog core based on characterized charge trap memory arrays, and evaluate its accuracy, energy efficiency, and area. Section \ref{sec:conclusion} concludes the paper.

\section{The Design Space of Analog Inference Architectures}
\label{sec:background}

\begin{figure}
    \centering
    \includegraphics[width=0.48\textwidth]{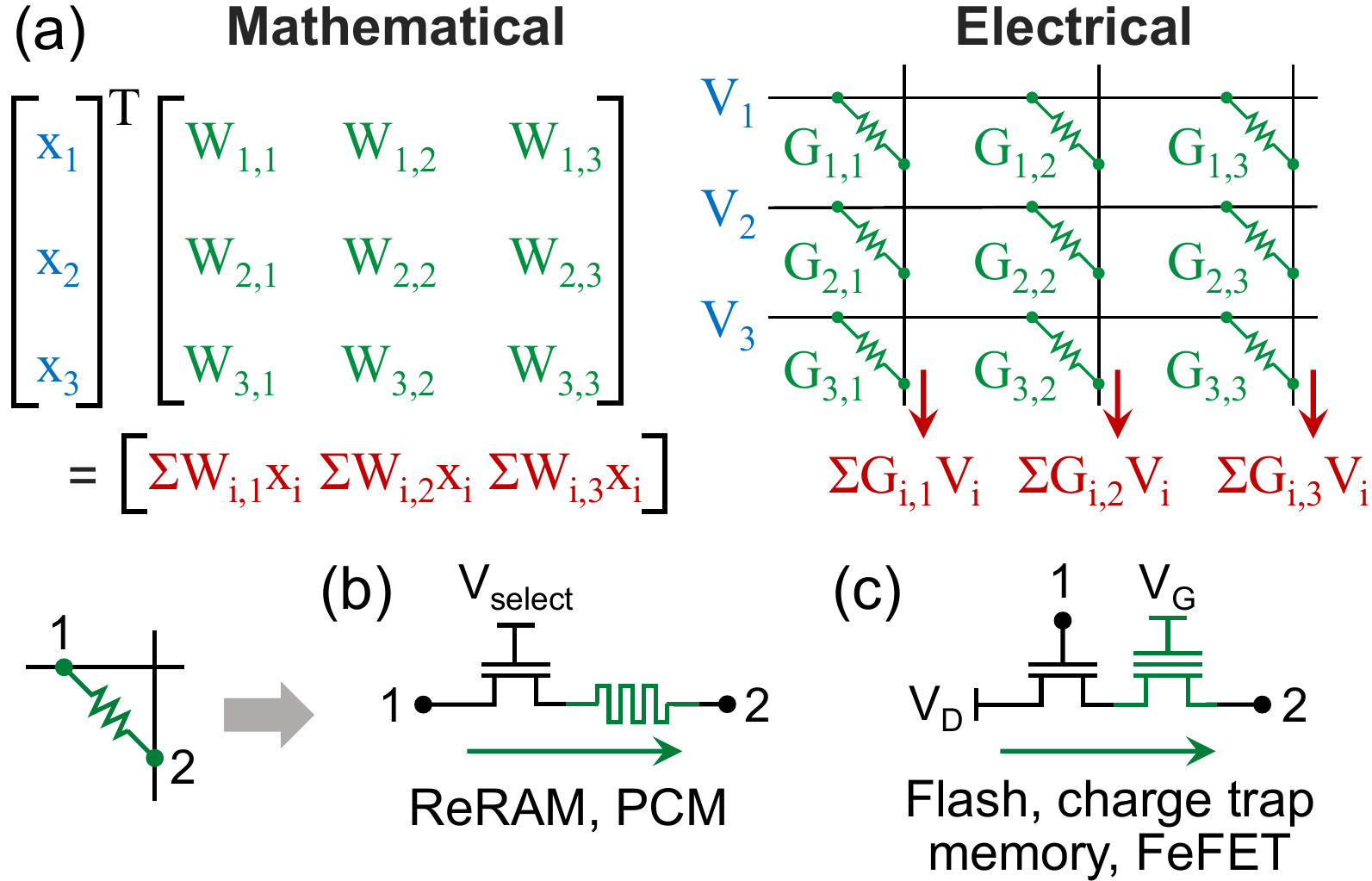}
    \caption{(a) Execution of an MVM $\vec{y} = \mat{W}\vec{x}$ within a memory array. (b)-(c) Two implementations of a memory cell that performs analog multiplication.}
    \label{fig:MVM}
  \vspace{-8pt}
\end{figure}

Analog accelerators perform matrix computations within the same memory arrays where the neural network weights are stored.
In contrast to digital architectures that spend significant energy to read operands from memory, \insitu{} computation eliminates the need to move weight data between processing elements. Within an array, individual analog MACs can also be conducted at a lower energy, higher density, and greater parallelism than digital MACs~\cite{Marinella18}.
Due to these potential advantages, \insitu{} MVM has attracted significant research attention for neural network inference~\cite{Shafiee16,Chi16,Nag18,Ankit19,Chou19,Li20}, as well as other applications~\cite{Bojnordi16,Song18,Feinberg18}.

Fig.~\ref{fig:MVM}(a) shows a conceptual example of an \insitu{} MVM array that computes $\vec{y}=\mat{W}\vec{x}$. The memory cell conductances $\mat{G}$ are set proportional to the values of $\mat{W}$, and the rows are driven by input voltages $\vec{V}$ that are proportional to $\vec{x}$. Each cell's current is an analog product of its conductance $G_{ij}$ and the applied voltage $V_i$. Kirchoff's law then accumulates these products on the bit line (column) current $I_j$ to form the dot product. The analog dot products are subsequently quantized using an ADC.

\Insitu{} MVM has been demonstrated using a wide variety of memory cell technologies~\cite{Tsai18,Sebastian20,Yu21}. Fig.~\ref{fig:MVM}(b) shows a 1T1R (1 transistor, 1 resistor) cell, which performs multiplication using Ohm's law across a two-terminal programmable resistor, such as a resistive random access memory (ReRAM) or phase change memory (PCM) device. During an MVM, the transistor is transparent. Fig.~\ref{fig:MVM}(c) shows an alternative cell design, more typically used with transistor-based memories such as flash memory~\cite{Guo17,Fick17,Agrawal20}, where a select transistor uses the input to gate the flow of current through the memory element (green).

The conceptual example in Fig.~\ref{fig:MVM} elides a number of practical implementation details. Prior work has proposed multiple approaches for data representation ($\mat{W}$ and $\vec{x}$) that differ from the mapping in Fig.~\ref{fig:MVM}. Table~\ref{tab:prior_accelerator} summarizes the design choices made by several recently proposed \insitu{} MVM accelerators, which are explained below. A recent review of analog inference accelerators can be found in Xiao \etal{}~\cite{Xiao20}.

\renewcommand{\arraystretch}{1.05}
\begin{table*}[t]
\caption{Comparison of data representation in selected prior work on analog \insitu{} inference accelerators}
\label{tab:prior_accelerator}
\begin{threeparttable}
        \begin{tabularx}{\textwidth}{|>{\centering\arraybackslash\hsize=2.4\hsize}X|>{\centering\arraybackslash\hsize=0.85\hsize}X|>{\centering\arraybackslash\hsize=0.64\hsize}X|>{\centering\arraybackslash\hsize=0.85\hsize}X|>{\centering\arraybackslash\hsize=0.95\hsize}X|>{\centering\arraybackslash\hsize=0.66\hsize}X|>{\centering\arraybackslash\hsize=0.65\hsize}X|}
        \hline
      \textbf{Accelerator} & \textbf{Bit slicing} & \textbf{Full \linebreak precision} & \textbf{Negative weights} & \textbf{\# rows used per MVM} & \textbf{ADC bits} $B_\text{ADC}$ & \textbf{DAC bits}\\
        \hline
      Genov \etal{}\cite{Genov00} & Yes, 1b/cell & No & One's comp & 128 & 6 & 1 \\
      \hline
      Memristive Boltzmann Machine\cite{Bojnordi16} & Yes, 1b/cell & Yes & Two's comp & 32 & 5 & 1 \\
      \hline
      ISAAC\cite{Shafiee16}, Newton\cite{Nag18} & Yes, 2b/cell & Yes & Offset & 128 & 8 & 1 \\
      \hline
      PUMA\cite{Ankit19} & Yes, 2b/cell & Yes & Offset & 128 & 8 & 1 \\
      \hline
      PRIME\cite{Chi16} & Yes, 4b/cell & No & Differential & 256 & 6 & 3 \\
      \hline
      Dot-Product Engine\cite{Hu16} & No & No & Offset & 128 & 4 & 4 \\
      \hline
      Sparse ReRAM Engine\cite{Yang19} & Yes, 2b/cell & Yes & Offset & 16 & 6 & 1 \\
      \hline
      CASCADE\cite{Chou19} & Yes, 1b/cell & Yes & -- & 64 & 10 & 1 \\
      \hline
      TIMELY\cite{Li20} & Yes, 4b/cell & No & -- & 256 & 8 & 8\tnote{*} \\
      \hline
      FORMS\cite{Yuan21} & Yes, 2b/cell & Yes & Retrain\tnote{$\dagger$} & 8 & 4 & 1 \\
      \hline
      Marinella \etal{}\cite{Marinella18} & No & No & Differential & 1024 & 8 & 1 \\
      \hline
      Joshi \etal{}\cite{Joshi20} & No & No & Differential & 512 & 8 & 8 \\
      \hline
      Yao \etal{}\cite{yao2020fully} (experimental) & No & No & Differential & 128 & 8 & 1 \\
      \hline
      Guo \etal{}\cite{Guo17} (experimental) & No & No & Differential & 784 & analog & 1 \\
      \hline
    \end{tabularx}
\begin{tablenotes}\footnotesize
\item[$\dagger$] The network is trained so that all weights on a column have the same sign. * Inputs are encoded in the temporal duration of a pulse.
\end{tablenotes}
\end{threeparttable}
  \vspace{-8pt}
\end{table*}

\subsection{Weight Bit Slicing vs Unsliced Weights}
\label{sec:background-weight-slicing}
To represent matrices with more bits than can be reliably programmed in a device, many systems use \emph{bit slicing}~\cite{Bojnordi16}.
In bit slicing, the bit representation of each matrix element is divided into multiple slices, and the results of bit sliced MVMs are combined via shift-and-add (S\&A) reduction~\cite{Shafiee16,Bojnordi16,Genov00}.
Equation~\ref{eq:bitslice} shows how a matrix of 6-bit integers can be divided into two slices of three bits each.
\begin{equation}
\label{eq:bitslice}
\left [\begin{matrix}
12 & 58 \\ 
 29 & 50 
\end{matrix} \right ] = 2^3
\left[ \begin{matrix}
1 & 7 \\ 
 3 & 6 
\end{matrix}\right ] + 
2^0
\left[ \begin{matrix}
4 & 2 \\ 
 5 & 2 
\end{matrix}\right ]
\end{equation}

\begin{figure}[t]
    \centering
    \includegraphics[width=0.46\textwidth]{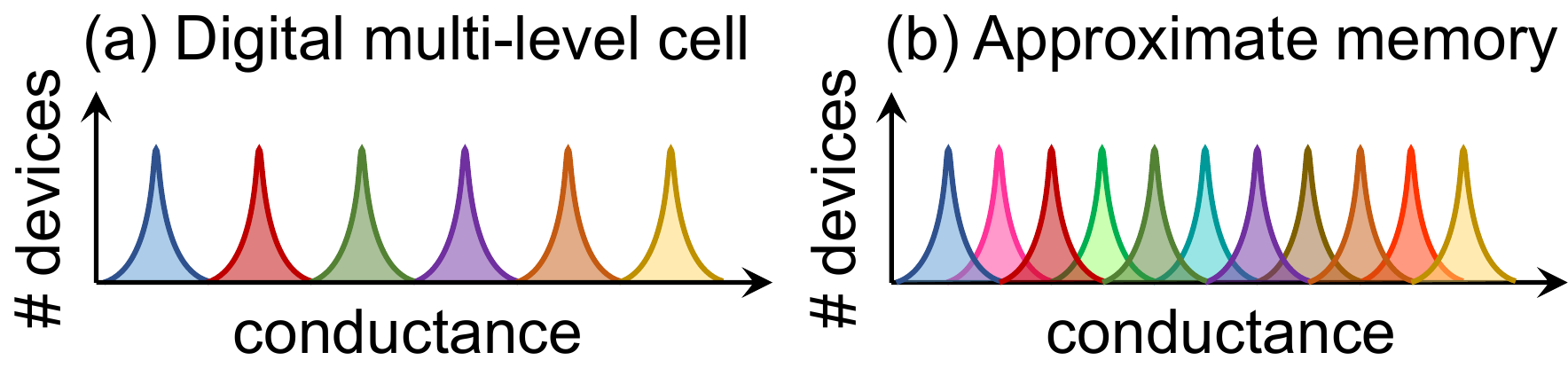}
    \caption{Conductance program distribution of a memory cell when used as two different types of memory.}
    \label{fig:approxMemory}
  \vspace{-8pt}
\end{figure}

Bit slicing admits the use of high-precision weights with more possible values than the number of programmable levels in a memory device. In particular, it allows the use of inherently binary memories such as SRAM that cannot otherwise implement multi-bit weights~\cite{Dong20}. Many accelerators use bit slicing as a way to tolerate analog memory cells with arbitrarily low precision, but this assumption has not been thoroughly evaluated on the basis of end-to-end inference accuracy and not just the weight precision.

To avoid the energy and area overheads of reading, digitizing and aggregating multiple bit-sliced arrays, the magnitude of a weight can also be fully encoded in one device~\cite{Joshi20,Hu16}. Unsliced weights ostensibly require very precise devices; however, for inference it can be sufficient to use analog memory cells not as multi-bit digital memories as in Fig.~\ref{fig:approxMemory}(a), but as \emph{approximate memories} shown in Fig.~\ref{fig:approxMemory}(b). The conductance state of an analog cell has a nonzero width due to process variations and noise. When used as a multi-bit digital memory, digital levels are mapped to states that have nearly zero overlap to enable statistically reliable readout of a single cell. When used as approximate memory, many more digital levels are mapped to the same conductance range by allowing states to overlap. The number of bits that can be mapped to a cell is ultimately limited by the resolution of the write circuitry and by the intrinsic physical resolution of the cell.

\subsection{Input Bit Slicing}
\label{sec:background-input-slicing}
Bit slicing can also be applied to multi-bit inputs; each analog MVM processes one slice of $\vec{x}$ and the full input is processed in multiple cycles. One-bit input slices are commonly used to avoid the high overhead of a multi-bit digital-to-analog conversion (DAC) per input on each MVM. A binary input voltage further allows the use of memory devices with a nonlinear $I$-$V$ curve, since ideally only two points along this curve are sampled\cite{Tsai18}. The cell configuration in Fig. \ref{fig:MVM}(c) also relies on one-bit input slices since the select gate functions as a binary switch.

Partial MVM results from multiple input bit slices can be aggregated digitally using a S\&A reduction network, similar to weight bit slices. The total number of ADC quantizations required per full MVM (all weight and input bits) is the product of the number of weight and input slices. Alternatively, S\&A accumulation of sequentially applied input bits can be conducted by switched-capacitor circuits prior to the ADC, so that only one quantization is needed for all input bit slices~\cite{Bavandpour20,Chou19,Ghodrati20}. Though feasible for 8-bit inputs, this technique cannot be scaled to arbitrarily many input bits due to the thermal noise floor on the analog signal.

\subsection{Handling Negative Numbers}
\label{sec:background-negative-numbers}

A variety of techniques have been developed to handle signed arithmetic with multi-bit weights and inputs. This work evaluates the two most common implementations of negative weights: \emph{offset subtraction} and \emph{differential cells}. These schemes will be described in more detail in Section \ref{subsec:weight_mapping}. Offset subtraction implements signed weights by including an offset in the conductance used to represent a zero. This then allows negative weights to be converted to positive conductances. This offset then needs to be subtracted from the dot product, either digitally or in the analog domain \cite{Shafiee16}.

In the differential cells scheme, a signed weight is represented using the difference in conductance of two memory cells. The specific implementation of the subtraction varies across designs, and can be performed in the analog domain or after digitization. Analog current subtraction can be executed using opposite-polarity voltage inputs and Kirchoff's law~\cite{Marinella18}, or within the bit line peripheral circuitry~\cite{Joshi20,yao2020fully,Guo17}. Though this scheme uses two cells per weight (or per weight slice) rather than one, it possesses some advantages over offset subtraction, as summarized in Section \ref{sec:key_ideas}.

Negative inputs can be handled by applying negative voltages to a resistive array~\cite{Marinella18}, or by using two differential pairs (four cells) per weight~\cite{Bavandpour19,Schlottmann11}. Notably, negative inputs are uncommon beyond the first layer of convolutional neural networks (CNNs) based on rectified linear (ReLU) activations. If both weights and inputs use one-bit slices, it is possible to use a two's complement representation for both~\cite{Bojnordi16}.

\subsection{The Full Precision Guarantee}
\label{sec:background-full-precision}

The conversion of an analog dot product to a digital result can incur a loss of precision. To provide theoretically digital accuracy from an analog MVM, prior work proposed the \emph{full precision guarantee} (FPG)~\cite{Shafiee16,Bojnordi16}. The FPG posits that if the ADC has a unique level for every possible output of the MVM operation, then there is no loss of information from digitization. The information content of the analog signal, equal to the number of bits needed to specify all possible dot product values, is a function of the operand widths and the number of summed terms~\cite{Shafiee16}:
\begin{align}
\label{eq:full_precision}
B_\text{out} = 
\begin{cases}
B_\text{W} + B_\text{in} + \log_2 N & \text{if} \, B_\text{W}>1, B_\text{in} > 1 \\
B_\text{W} + B_\text{in} + \log_2 N -1  & \text{otherwise} \\
\end{cases}
\end{align}
where $B_\text{W}$ is the number of weight bits per cell, $B_\text{in}$ is the number of input bits per ADC operation, and $N$ is the number of rows activated in an MVM. Notably, if input bit slices are accumulated digitally, $B_\text{in}$ is the number of input bits per slice; if they are accumulated by analog circuitry, $B_\text{in}$ is the smaller of the full input resolution or the circuit's resolution. A non-integer value of $B_\text{out}$ simply means that the number of possible MVM outputs is not a power of two.

The FPG can be stated as:
\begin{align}
B_\text{ADC} = \lceil B_\text{out} \rceil
\end{align}
where $B_\text{ADC}$ is the ADC effective number of bits. Typically, $B_\text{in}$, $B_\text{W}$ and $N$ are chosen such that $B_\text{out} \approx 8$ bits. Since the ADC cost rapidly becomes prohibitive with resolution~\cite{Murmann20}, prior work using the FPG has been limited to smaller arrays and/or fewer bits per weight. Shafiee \etal{}~\cite{Shafiee16} proposed a `flipped' encoding of weight values to reduce the required ADC resolution to $\lceil B_\text{out} \rceil -1$ bits.

In many systems, the final result after aggregating all slices is truncated before being passed to the next layer. This means that not all $B_\text{out}$ bits from every slice are useful. Prior work has proposed avoiding this wasted computation by dynamically tuning the ADC resolution on a slice-wise basis~\cite{Nag18,Chi16}.

As shown in Table \ref{tab:prior_accelerator}, not all systems adopt the FPG. This design choice will be evaluated in Sections \ref{sec:key_ideas} and \ref{sec:ADC}.

\subsection{Direct Weight Transfer vs Retraining}
\label{sec:background-retraining}
To improve inference accuracy in the presence of analog errors, many methods have been proposed to integrate these errors into the training process.
A common approach is to add noise to weights and activations during forward propagation~\cite{Joshi20,Klachko19,He19,Jain19,Long19_2}.
However, these techniques incur additional training overhead and are potentially difficult to co-integrate with state-of-the-art training workflows.
Therefore, this paper focuses on accuracy with direct weight transfer: weights are mapped as-is to the memory cells, without any retraining or compensation for errors post-training~\cite{Hu16,Jain19}.
This work performs some pre-processing to calibrate the ADC limits, which is similar to the process used to quantize neural networks for digital inference\cite{Jacob18}.
This matches the standard by which the accuracy of digital accelerators is evaluated~\cite{Reddi20}.

\section{Design principles for an error-resilient analog accelerator}
\label{sec:key_ideas}

This section summarizes the key design principles that enable an error-resilient analog inference accelerator. The remaining sections provide the modeling methodology and the results that support these general conclusions.

\subsection{Precise Weight Representation $\neq$ Precise Dot Product Computation}
\label{subsec:bit_slicing}

\begin{figure}[t]
    \centering
    \includegraphics[width=0.25\textwidth]{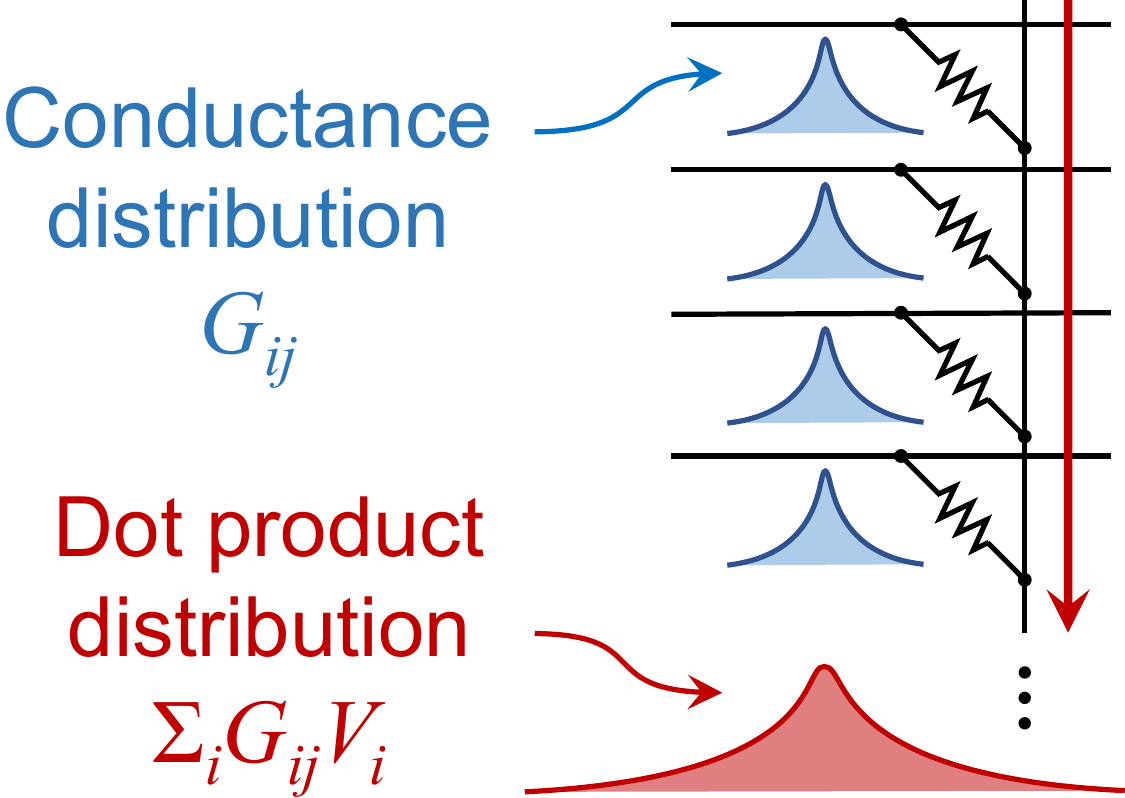}
    \caption{Accumulation of cell errors along a bit line.
}
    \label{fig:accumulation}
   \vspace{-6pt}
\end{figure}

Bit slicing allows weights to be represented with arbitrarily high precision using memory cells that have only a few reliably distinguishable conductance states due to programming errors. However, the most important consequence of cell errors on accuracy is not their effect on the fidelity of individual weights, but rather the effect of summed cell errors on the fidelity of dot products, since these are the quantities that propagate from layer to layer during inference. This is illustrated in Fig.~\ref{fig:accumulation}. Each cell has a random deviation $\Delta G_{ij}$ from its target conductance, and these errors are added when currents from multiple cells are summed on a bit line. If $\langle\Delta I_{ij}\rangle$ is the expected error in the product $G_{ij}V_i$ that results from this conductance error, then the expected error in the dot product $I_j$ accumulated on the bit line is:
\begin{align}
\label{eqref:error_accum}
\langle \Delta I_j \rangle^2 = N \langle \Delta I_{ij} \rangle^2
\end{align}
Some prior work uses the dot product error $\langle \Delta I_j \rangle$ as a starting point in an accuracy analysis~\cite{Sze19,Chou19,Rekhi19}, but this obscures the design choices that affect the size of the accumulated error.

Importantly, the above equation holds whether or not the distributions of conductance states within a cell overlap. To minimize the error in the dot product, the absolute width of the distribution $\Delta G$ is more important, and this quantity is not improved by ensuring that the utilized states are well separated. Therefore, \emph{bit slicing does not provide a fundamental advantage to accuracy compared to approximate memories}, and conversely, it cannot be relied upon to save the accuracy when memory cells with inherently large errors are used. Many of the works listed in Table \ref{tab:prior_accelerator} choose to use bit slicing without evaluating the accuracy with unsliced weights, with the implicit assumption that the accuracy would fall significantly without bit slicing. In Section \ref{sec:errors}, we show that bit slicing does not in fact provide a large advantage to accuracy for the same device conductance precision.

Bit slicing can nonetheless provide a \emph{small} improvement to accuracy, as will be explained in Sections \ref{subsec:ind_error} and \ref{subsec:proportional_error}. The origin of this benefit is subtle and does not stem from having well-separated memory states. Since the benefit tends to be small, it must be considered carefully against the large energy and area overheads of bit slicing, as shown in Section \ref{sec:sonos}.

If the accumulated error $\Delta I_j$ can be reduced below the least significant bit (LSB) of the ADC, its propagation to the next layer can be suppressed. This can be achieved by using smaller arrays\cite{Yang19,Lin18}, but this is inefficient as it amortizes the ADC energy cost over fewer MACs. Error correcting codes can correct a fraction of the dot product errors\cite{Feinberg18_2}, but the simplest and least costly method of reducing these errors is to proportionally map weights to conductances.

\begin{figure}[t]
    \centering
    \includegraphics[width=0.4\textwidth]{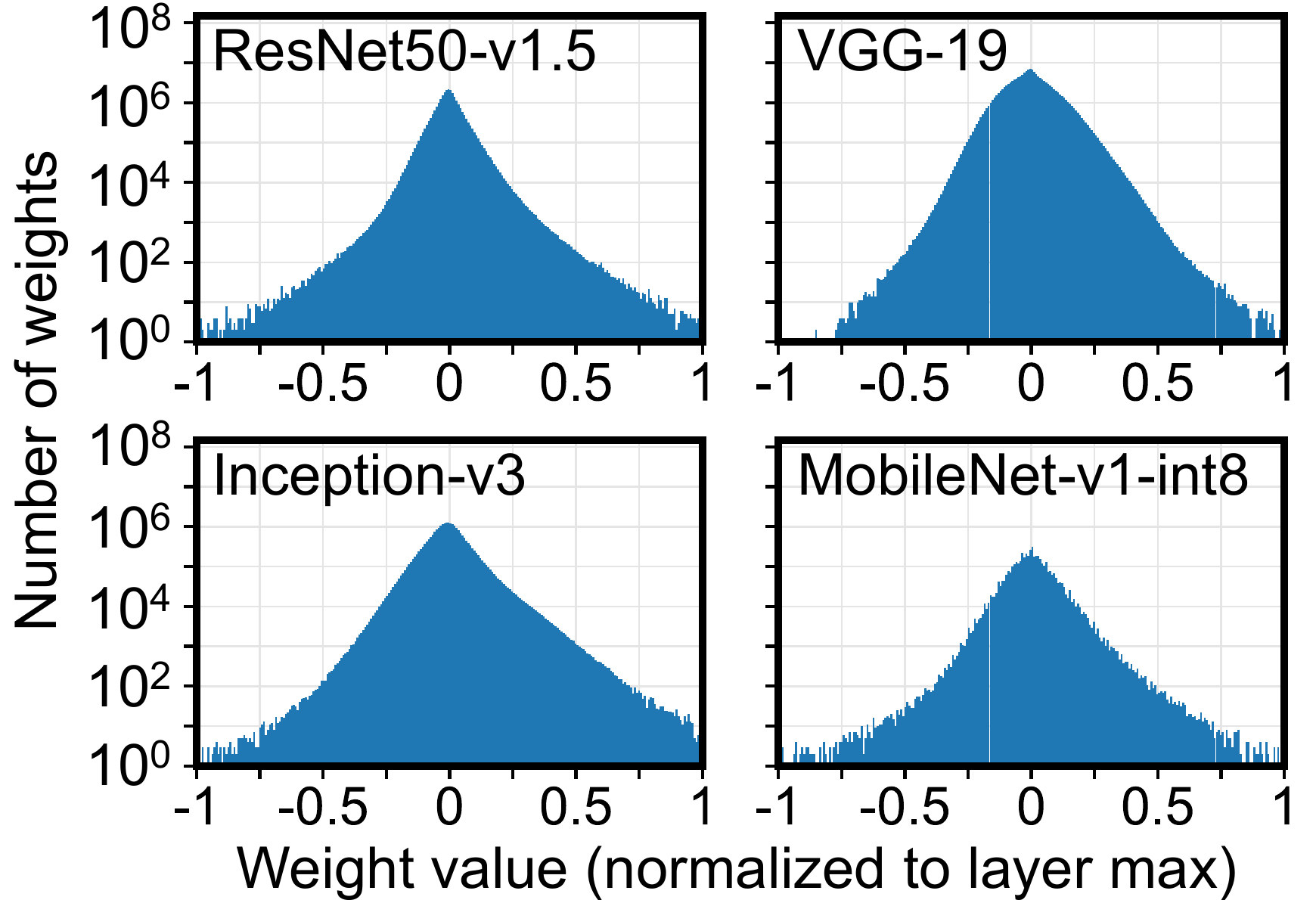}
    \caption{Weight value distributions of several popular ImageNet neural networks.
}
    \label{fig:weight_dists}
   \vspace{-6pt}
\end{figure}

\subsection{Proportional Mapping Reduces Errors}
\label{subsec:proportionality}
\subsubsection{Weight Proportionality}

A very common property of neural networks is the abundance of low-valued or zero-valued weights. This is illustrated in the weight value distributions shown in Fig.~\ref{fig:weight_dists} of four popular ImageNet neural networks. In digital inference accelerators, this property can be exploited to greatly compress the network size (via pruning) and the resultant sparsity can be used to save computation\cite{Han15,Han16EIE}. Pruning is more difficult to exploit in analog accelerators, due to the rigid structure of a memory crossbar\cite{Xiao20}. Nonetheless, it is possible to exploit zero and small-valued weights in analog accelerators by using \textbf{proportional mapping}: a linear relationship between numerical values in the algorithm and the physical quantities in the analog hardware.

With proportional mapping, weight values are mapped to conductances in proportion to their magnitude. This is implemented by using differential cells to encode negative weights in the manner described in Section \ref{sec:background-negative-numbers}, and by using cells with high On/Off ratio ($G_\text{max}/G_\text{min}$). Together with the strongly zero-peaked weight distributions in neural networks, proportional mapping can reduce the average cell conductance by orders of magnitude, as shown in Section \ref{subsec:imagenet}.

Reduction of the average conductance is important because two types of analog errors tend to increase proportionally with conductance or current. First, the cell programming error $\Delta G$ typically increases with the programmed conductance $G$, as will be described in Section \ref{sec:errors}. For some technologies, like flash memory, this is a fundamental property of the device. Another source of error that increases with cell conductance is parasitic voltage drops across the columns and/or rows of the array, which nonuniformly distort the elements of a weight matrix as described in Section \ref{sec:parasitics}. Proportional mapping mitigates both of these errors, by matching the least-error conductance states to the most-used weight values.

\subsubsection{Dot Product Proportionality}

Proportional mapping is also important between dot products and analog outputs. Neural networks natively possess some tolerance to low-resolution activations during inference. Activations in ImageNet neural networks, for example, can typically be quantized to 8 bits after training without losing significant inference accuracy\cite{Jacob18}. Can analog systems exploit this to perform accurate ImageNet inference with no more than 8 bits of ADC resolution? The answer is yes, and the key enabler is \emph{dot product proportionality}. While an activation may tolerate quantization to 8 bits, this property might be lost if the same information is encoded in a quantity that is not proportional to the original activation. Ensuring proportionality between analog outputs and dot products connects the ADC resolution requirement to the algorithm's intrinsic precision requirements. Dot-product proportionality largely follows from weight proportionality, with the requirement that the current subtraction in differential cells be conducted in analog. This will be explored in Section \ref{sec:ADC}.

\subsection{The Full Precision Fallacy}
\label{subsec:FPG_fallacy}

The FPG requires the ADC to have a unique level for every possible output of an analog MVM, and thus match the precision of a digital processor. To be compatible with practical ADC resolutions ($\sim$8 bits), the FPG bounds the amount of computation that can be executed in the analog domain before digitization. This is expressed by Equation~\eqref{eq:full_precision}. There are two fundamental problems with the FPG.

First, the FPG is only meaningful if the accumulated cell error on all bit lines is below the LSB of the ADC. When cell errors are present, the analog input to the ADC does not necessarily have $B_\text{out}$ bits of precision as given by Equation \eqref{eq:full_precision}. Thus, in practice, satisfying the FPG requires not only the correct ADC resolution but also sufficiently accurate memory cells to ensure that the ADC resolution is fully utilized. Some early work on \insitu{} MVM explicitly set the ADC resolution to match the expected level of accumulated cell error, using fewer bits than required by the FPG~\cite{Genov00,Rekhi19}.

Second, the FPG is imposed at the level of the analog MVM kernel, typically without full consideration of its utility for the accuracy of neural network inference. By focusing on the precision of an individual kernel rather than end-to-end system requirements, the FPG creates inefficiencies, as predicted by the end-to-end argument of Saltzer \etal{}~\cite{Saltzer84}. Specifically, in Section \ref{sec:ADC}, we show that in systems with dot-product proportionality, the FPG is too conservative. In these systems, \emph{the ADC resolution requirement can be decoupled from the hardware configuration and dictated instead by the end-to-end accuracy of the neural network application}.  Fortuitously, the resolution requirement of ImageNet neural networks is also $\sim$8 bits\cite{Jacob18}. Removing the constraints of the FPG enables much more analog computation to be done for the same ADC resolution, improving energy efficiency.

\section{Accuracy Evaluation Method}
\label{sec:methods}

\begin{figure*}[t]
    \centering
    \includegraphics[width=\textwidth]{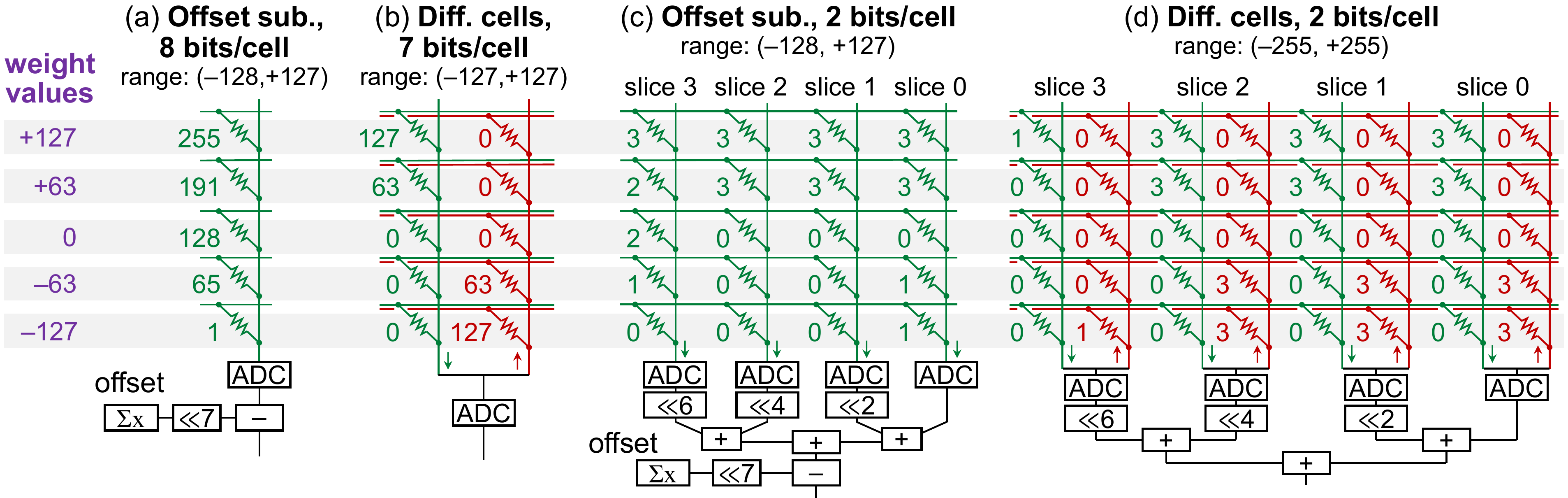}
    \caption{Four schemes for mapping weight values to conductance values in a memory array.
}
    \label{fig:array_setup}
   \vspace{-8pt}
\end{figure*}

This section describes the methodology for inference accuracy evaluation for the results presented in the remaining sections. Unless otherwise stated, neural networks are quantized to 8-bit precision, a common use case for inference~\cite{Jacob18,Jouppi17}.

\subsection{Mapping Weights to Conductances}
\label{subsec:weight_mapping}

We assume a simple, parameter-less procedure to map a layer's weights onto device conductances. First, the floating-point weight matrix $\mat{W}_\text{FP}$ is scaled by the maximum absolute value $\text{max}|\mat{W}_\text{FP}|$ into the range $[-1,+1]$. To quantize the weights to 8 bits, these weights are further scaled to the range $[-127,+127]$, then rounded to integers in this range. The same process is used for the weights of all layers.

The quantized weight matrix $\mat{W}$, with integer values in the range $[-127,+127]$, is then decomposed into one or more non-negative integer-valued matrices that can be mapped to conductances. The specific decomposition depends on the method used to handle negative weights (offset subtraction vs. differential cells) and encode weight precision (with or without bit slicing). Several examples are shown in Fig.~\ref{fig:array_setup} for an 8-bit matrix. Although these representations are functionally equivalent in the absence of analog errors, they differ greatly in their sensitivity to these errors. These methods represent the majority of proposed analog accelerators.

\subsubsection{Mapping without Bit Slicing}

Fig.~\ref{fig:array_setup}(a) shows the 8-bit matrix $\mat{W}$ is mapped using offset subtraction without bit slicing, following the equation:
\begin{align}
\label{eq:offset}
\mat{W}\vec{x} = \mat{W}_\text{prog}\vec{x} - 2^7\,\mat{I}\vec{x}
\end{align}
where $\mat{I}$ is the identity matrix and $\mat{W}_\text{prog}$ is a strictly non-negative 8-bit matrix in the range [1,~255] that can be mapped directly onto conductances. This matrix has an offset such that a zero weight in $\mat{W}$ is mapped to a value of $2^7$ in $\mat{W}_\text{prog}$. This offset is subtracted from the MVM result to represent negative weights. Computing the offset term requires summing the elements of $\vec{x}$, which can be done digitally. Shafiee \etal{}\cite{Shafiee16} also proposed an analog computation of this offset with a ``unit column'', which will be evaluated in Section \ref{subsec:ind_error}. We note that a value of $-128$ in $\mat{W}$ can be mapped by a value of 0 in $\mat{W}_\text{prog}$, but this state is left unused. 

Fig.~\ref{fig:array_setup}(b) shows how the same matrix is mapped using differential cells, following the equation:
\begin{align}
\label{eq:diff}
\mat{W}\vec{x} = \mat{W}^+\vec{x} - \mat{W}^-\vec{x}
\end{align}
where the strictly positive weight matrices $\mat{W}^+$ and $\mat{W}^-$ have 7-bit values in the range [0,~127], and are programmed onto the conductances of two sets of memory cells. This definition leaves some ambiguity about how two conductances are decided from a single weight value. This paper evaluates the convention where one cell in the pair encodes the magnitude of positive weights, while the other encodes the magnitude of negative weights. This means that at least one cell in every pair is left in the lowest conductance state. Note that the weight magnitudes in this scheme are directly mapped to the conductances of 7-bit cells, ensuring a proportional mapping.

The integer values in the non-negative matrices on the right sides of Equations \eqref{eq:offset} and \eqref{eq:diff} are mapped linearly to conductances. A value of 0 is mapped to the minimum conductance state $G_\text{min}$, while the maximum value in the range is mapped to $G_\text{max}$. Intermediate integers are linearly mapped to intermediate conductances.

\subsubsection{Mapping with Bit Slicing}

Fig. \ref{fig:array_setup}(c) shows an example of offset subtraction with bit slicing, which implements the following mapping:
\begin{align}
\label{eq:bitslicing_offset}
\mat{W}\vec{x} = 2^6 \mat{W}_3\vec{x} + 2^4 \mat{W}_2\vec{x} + 2^2 \mat{W}_1\vec{x} + \mat{W}_0\vec{x} - 2^7\mat{I}\vec{x}
\end{align}
where $\mat{W}_i$ are the 2-bit slices of $\mat{W}$ from lowest to highest significance. Each element of $\mat{W}_i$ is integer-valued in the range [0,~3] and mapped to the conductance of a single cell. The offset is subtracted after the results of the slices are aggregated. As in the non-bitsliced case, this is not a proportional mapping, since a zero weight is mapped to an intermediate conductance in the top slice.

Fig. \ref{fig:array_setup}(d) shows an example of differential cells with bit slicing, which implements the following mapping:
\begin{align}
\label{eq:bitslicing_diff}
\begin{split}
\mat{W}\vec{x} = 2^6 \left(\mat{W}_3^+\vec{x} - \mat{W}_3^-\vec{x} \right) + 2^4 \left(\mat{W}_2^+\vec{x} - \mat{W}_2^-\vec{x} \right) \\+ 2^2 \left(\mat{W}_1^+\vec{x} - \mat{W}_1^-\vec{x} \right) + \left(\mat{W}_0^+\vec{x} - \mat{W}_0^-\vec{x} \right)
\end{split}
\end{align}
where each 2-bit matrix $\mat{W}_i^\pm$ is integer-valued in the range [0,~3]. This method uses a sign-magnitude representation and slices the magnitude bits across multiple cells. Within a slice, the magnitudes of positive weights are mapped to $\mat{W}_i^+$ and the magnitudes of negative weights are mapped to $\mat{W}_i^-$, and the resulting bit line currents are subtracted. The most significant slice is proportional to the weight value, and a zero weight is mapped entirely onto the `0' state in all slices, as shown in Fig. \ref{fig:array_setup}(d). Because the four slices together represent 8 magnitude bits, the hardware in Fig.~\ref{fig:array_setup}(d) can map a 9-bit signed weight in the range [$-255,+255$].

\subsubsection{Input Bit Accumulation}
\label{subsec:input_bits}

For all of the schemes above, one-bit input slices are assumed to simplify the input DAC and device requirements. For differential cells, results from different input bits are sequentially accumulated using analog circuitry as described in Section \ref{sec:background-input-slicing}, such that $B_\text{in}=8$ bits. For offset subtraction, analog accumulation requires summing all of the 8-bit elements of the input vector $\vec{x}$ to compute the offset, which is more complex than summing the elements of $\vec{x}$ one bit at a time. To avoid this overhead and to provide a baseline that is similar to prior work \cite{Shafiee16,Nag18,Ankit19}, offset subtraction is evaluated with digital S\&A accumulation of input bits ($B_\text{in} = 1$ bit).

\subsection{Accuracy Simulation of Analog MVMs}

For a realistic accuracy simulation of an analog inference accelerator, we extend CrossSim~\cite{CrossSim} with a highly parameterizable model for an analog MVM array. CrossSim imports a Keras neural network model~\cite{Keras} and maps the weight matrix of each convolution and fully-connected layer to one or more memory arrays, representing different bit slices and matrix partitions, according to a chosen mapping scheme. Every analog MAC is simulated during inference. Digital operations such as the S\&A aggregation of weight slices, ReLU activation, and inter-layer communication are assumed to be error-free. Convolutions are unrolled into a sequence of sliding window MVMs, executed on arrays of size $K_x K_y N_\text{ic} \times N_\text{oc}$ as described by Shafiee \etal{}~\cite{Shafiee16} ($K_x \times K_y$ is the 2D filter size, $N_\text{ic}$ and $N_\text{oc}$ are input and output channel dimensions).

The modeling of random cell programming errors, ADC quantization, and parasitic voltage drops are described in Sections~\ref{sec:errors}, \ref{sec:ADC}, and \ref{sec:parasitics}, respectively, where the accuracy impact of each non-ideality is analyzed separately. This work does not study the effect of cycle-to-cycle read noise, which is similar to that of programming errors. Read noise has a weaker effect than programming errors when input bit slicing is used, as explained in Section~\ref{subsec:ind_error}. This work also does not study conductance drift over time, which is less generalizable across technologies.

The non-idealities mentioned above grow in severity with the number of cell currents summed on the same bit line, and hence limit the number of rows in the array. The maximum array size is treated as a parameter, and matrices that require more rows are partitioned evenly across equally sized arrays. The results from each array are separately digitized (and possibly clipped) before they are added.

To reduce digital processing overheads, batch normalization parameters are folded into the weight matrix of a convolution for all the evaluated networks~\cite{Jacob18}. Since the bias weights can lie in a different range from the other weights, representing them together in the same array can cause a loss of precision~\cite{Jacob18}. Therefore, for all layers the bias weights are stored separately from the array and added digitally to the MVM results.

\subsection{ImageNet Neural Network Benchmark}
\label{subsec:imagenet}

\begin{figure}[t]
    \centering
    \includegraphics[width=0.46\textwidth]{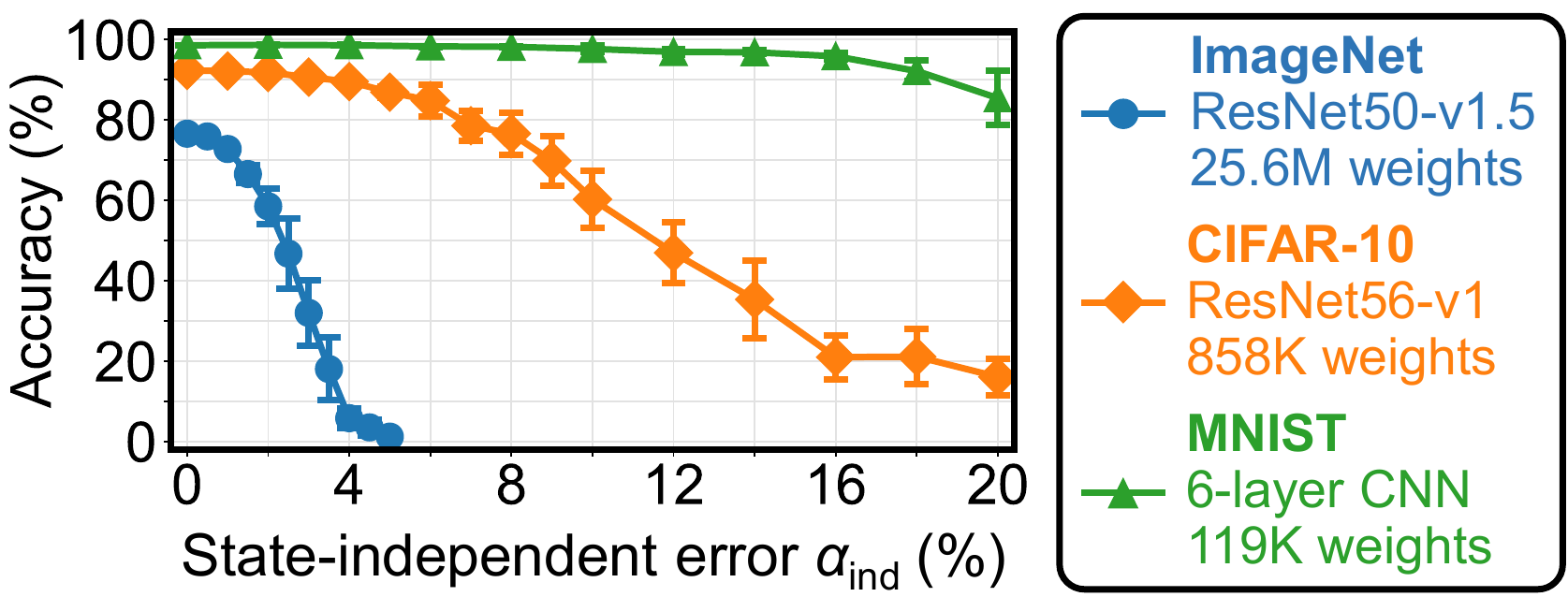}
    \caption{Accuracy sensitivity to state-independent cell errors for networks trained on three datasets. All cases assume unsliced weights with differential 7-bit cells.
}
    \label{fig:tasks}
   \vspace{-8pt}
\end{figure}

Fig.~\ref{fig:tasks} highlights the importance of using a realistic dataset for accuracy evaluations. The sensitivity to cell errors (described more fully in Section~\ref{sec:errors}) differs dramatically for networks trained on three datasets---ImageNet~\cite{Deng09}, CIFAR-10~\cite{cifar10}, and MNIST\cite{mnist}---with ImageNet being by far the most sensitive. The validity of any study on accuracy in analog accelerators is therefore bounded by the complexity of the inference task.

To emulate a realistic machine learning application, most of the accuracy evaluation in this paper is based on the ResNet50-v1.5 network for ImageNet, using the reference implementation from the MLPerf Inference Benchmark v0.5~\cite{Reddi20,MLPerfGit}. To compare the error sensitivity of different neural networks (Section \ref{subsec:networks}), we include three other popular ImageNet models: VGG-19\cite{Simonyan14}, Inception-v3\cite{Szegedy16}, and MobileNet-v1\cite{Howard17}. For VGG-19 and Inception-v3, we use the reference implementations included in Keras Applications\cite{Keras}. For MobileNet-v1, we evaluate the quantized implementation with 8-bit integer weights that is provided as part of the MLPerf Inference Benchmark v0.5\cite{MLPerfGit}. The MobileNet model is quantized to 8 bits during training, since the same model trained without quantization loses significant accuracy when quantized after training\cite{Reddi20}. Finally, we include a version of ResNet50-v1.5 that was trained by Nvidia at 4-bit precision, which will be described in Section \ref{sec:quant}. Table \ref{tab:networks} shows the accuracy of the evaluated networks on the ImageNet validation set, before applying any errors. The accuracy is also shown on a fixed subset of 1000 images, which is used for the sensitivity analyses in this paper for computational tractability.

\renewcommand{\arraystretch}{1.05}
\begin{table}[t]
\caption{Evaluated neural networks}
\vspace{3pt}
\label{tab:networks}
\setlength\tabcolsep{0pt}
    \begin{tabularx}{0.49\textwidth}{|>{\centering\arraybackslash\hsize=1.45\hsize}X|>{\centering\arraybackslash\hsize=0.65\hsize}X|>{\centering\arraybackslash\hsize=1.05\hsize}X|>{\centering\arraybackslash\hsize=0.85\hsize}X|}
        \hline

       \centering \multirow{2}{*}{\textbf{Neural network}} & \centering \multirow{2}{*}{\shortstack{\textbf{\#}\\\textbf{weights}}} & \multicolumn{2}{c|}{\textbf{ImageNet top-1 accuracy}} \\
    \cline{3-4}
      & & \centering 50,000 images &  \centering\arraybackslash 1000 images  \\ 
        \hline
      ResNet50-v1.5 & 25.6M & 76.466\% & 77.5\% \\
      \hline
      Inception-v3 & 23.9M & 77.876\% & 77.8\% \\
      \hline
      VGG-19 & 143.7M & 71.256\% & 70.2\% \\
      \hline
      MobileNet-v1 (int8) & 4.3M & 70.614\% & 71.8\% \\
      \hline
      ResNet50-v1.5 (int4) & 25.6M & 76.154\% & 76.6\% \\
      \hline
    \end{tabularx}
  \vspace{-8pt}
\end{table}

Weights are quantized to 8 bits before being mapped to hardware, and activations are quantized to 8 bits during inference. Except in the case of MobileNet, deployment at 8-bit precision does not need retraining, but the numerical range of the activations must be optimized to reduce quantization and clipping errors~\cite{Jacob18}. This is done by first running the model at floating-point precision and saving the activation values for all layers, using the MLPerf Inference calibration subset of 500 images\cite{MLPerfCalibrationGit}. For each layer's collected activations $\vec{x}$, the range $(x_\text{min},x_\text{max})$ is found that minimizes the L1-norm error $\epsilon = ||\vec{x}-\vec{x}_Q||_1$, where $\vec{x}_Q$ is obtained by clipping and quantizing $\vec{x}$ to $M$ bits in this range. A value of $M=12$ was found to yield an optimal inference accuracy for 8-bit activations. The value of $M$ does not correspond to any physical quantity in the system, and differs from the activation resolution because $\epsilon$ is not a true proxy for inference accuracy. The same activation ranges are used for all hardware implementations of a neural network. The resolution and range of the activations do not directly correspond to those for the ADCs, since multiple digitizations, a bias, and an activation function may be needed to produce one output activation. The ADC ranges are discussed in Section \ref{sec:ADC}.

\begin{figure}[t]
    \centering
    \includegraphics[width=0.48\textwidth]{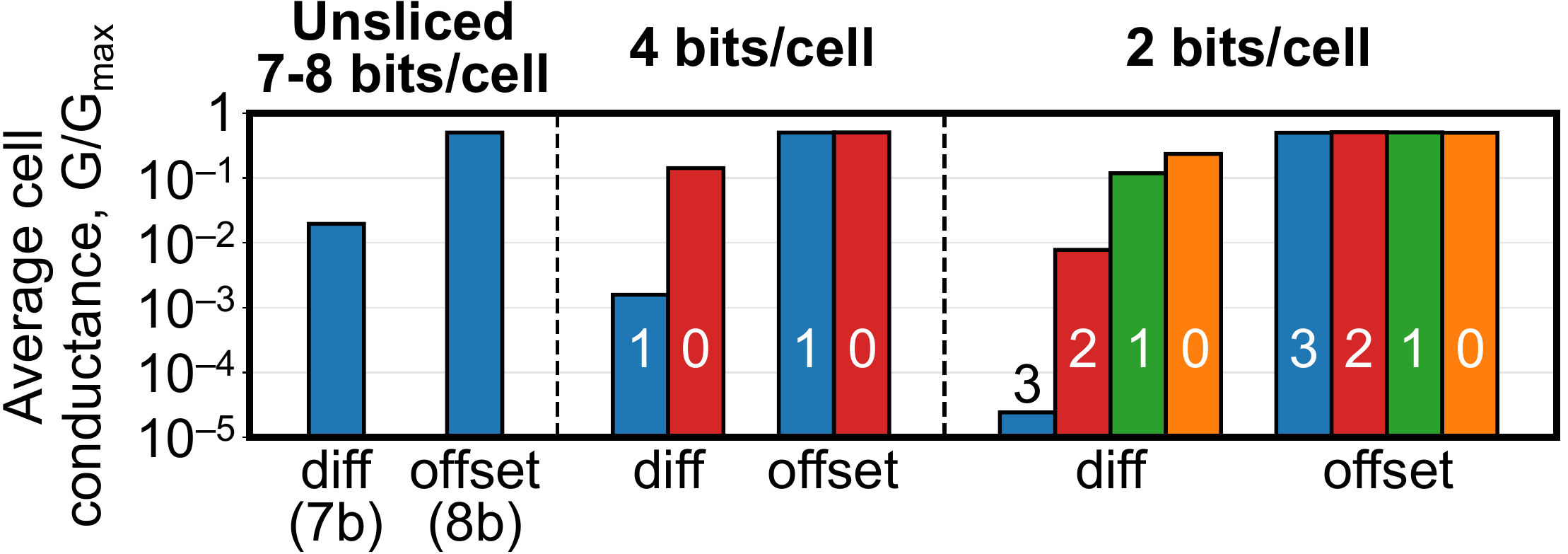}
    \caption{Average cell conductances for several different schemes to map the 8-bit weights in ResNet50-v1.5. The bars are labeled by bit slice (0 is lowest). 
}
    \label{fig:conductance_stats}
   \vspace{-10pt}
\end{figure}

Fig.~\ref{fig:conductance_stats} shows the average conductance in each bit slice for the data mapping schemes in Fig.~\ref{fig:array_setup} when implementing ResNet50-v1.5. Here, an infinite On/Off ratio ($G_\text{min} = 0$) is assumed.
Fig.~\ref{fig:conductance_stats} shows that \emph{using differential cells reduces the average cell conductance by multiple orders of magnitude} in the case of unsliced weights and in the higher slices with bit slicing. In the lower bit slices or when using offset subtraction, the average cell conductance is close to $50\%$ of $G_\text{max}$. As described in Section \ref{sec:key_ideas}, this conductance reduction is a consequence of proportional mapping and the abundance of low-valued weights in the neural network. The following sections will explore the implications of the conductance distribution on inference accuracy.

\section{Robustness to Cell Errors}
\label{sec:errors}

Due to process variations and device and circuit limitations, there is always some uncertainty in the conductance of a programmed cell. This section considers the effect of cell conductance errors on end-to-end inference accuracy. Except in Section \ref{subsec:networks}, all results are based on ResNet50-v1.5.

\subsection{Error Properties of Memory Devices}
\label{subsec:devices}

\begin{figure}[t]
    \centering
    \includegraphics[width=0.49\textwidth]{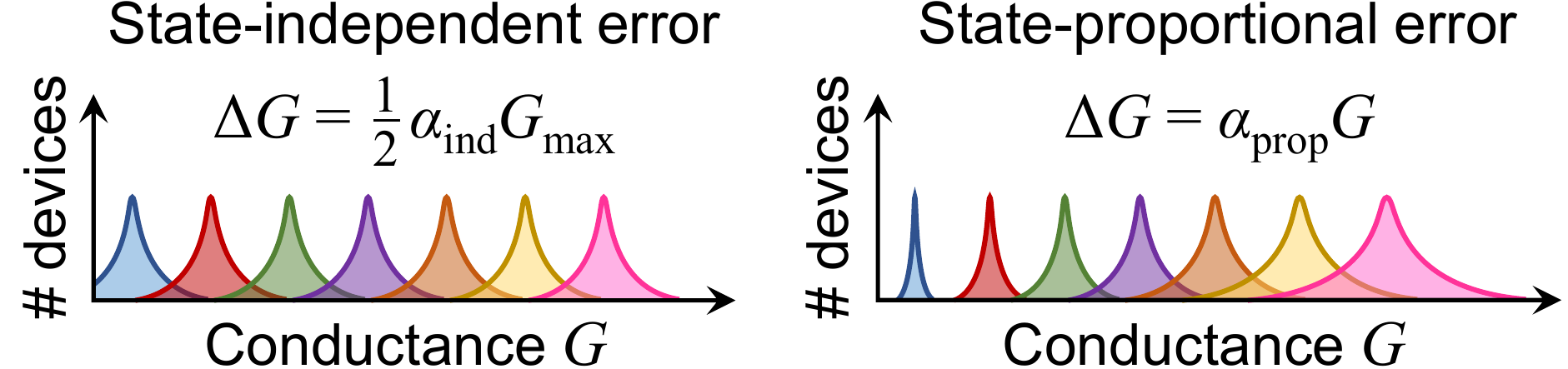}
    \caption{Two models for cell conductance error.
}
    \label{fig:error_types}
   \vspace{-8pt}
\end{figure}

Fig.~\ref{fig:error_types} depicts two simple models of conductance error in memory devices. In the state-independent error model, the expected error $\Delta G$ does not depend on the conductance $G$ and can be expressed as a fixed fraction of $G_\text{max}$. In the state-proportional error model, $\Delta G$ is proportional to the conductance; a smaller conductance has a smaller error. The parameters $\alpha_\text{ind}$ and $\alpha_\text{prop}$ are defined such that when the two are equal, the corresponding errors $\Delta G$ are the same at the midpoint conductance: $G = 0.5G_\text{max}$.

While real devices cannot be perfectly described by these models, many memory devices have the property that $\Delta G$ increases with $G$. As will be explained in Section \ref{sec:sonos}, flash memory has approximately state-proportional error properties when operated in the subthreshold regime; this is due to the exponential dependence of source-drain conductance on the amount of stored charge. The property that the error $\Delta G$ increases with $G$ has also been seen in PCM~\cite{Joshi20} and some ReRAM devices~\cite{Hayakawa15}. In these cases, $\Delta G$ is not strictly proportional to $G$, so the behavior is a mixture of the two error models analyzed here. Some other ReRAM devices have properties that are closer to state-independent error\cite{yao2020fully,Milo21}. Section \ref{sec:sonos} will evaluate the accuracy of a real memory device with a more complex state-dependent error characteristic.

To model cell errors, the conductance $G$ of every cell in the network is perturbed with an error that is sampled from a normal distribution. The distribution has zero mean and a standard deviation $\Delta G$, based on the equations in Fig. \ref{fig:error_types}. These perturbed conductances are then used to simulate inference on 1000 images. This process is repeated ten times, with re-sampled cell errors, to obtain the variance in accuracy over these images.

To study the effect of cell errors alone, this section does not include ADC quantization. Without ADCs, the accuracy is independent of array size as cell errors are allowed to accumulate over the full size of the weight matrix (up to 4608 inputs in ResNet50-v1.5). Thus, the results here represent the worst-case effect of cell errors. The sensitivity to cell errors in the presence of an ADC will be shown in Sections \ref{sec:quant} and \ref{sec:sonos}.

Unless otherwise noted, the following evaluation assumes that $G_\text{min} = 0$. The conductance On/Off ratio needed to approximate this idealization is found in Section \ref{subsec:proportional_error}. For differential cells, the examples with bit slicing use 9-bit weights to fully utilize the representational range of the hardware, while the unsliced case uses 8-bit weights. The difference in accuracy between 8-bit and 9-bit weights is 0.6\% on this subset without cell errors.

\subsection{Sensitivity to State-Independent Errors}
\label{subsec:ind_error}

\begin{figure}[t]
    \centering
    \includegraphics[width=0.48\textwidth]{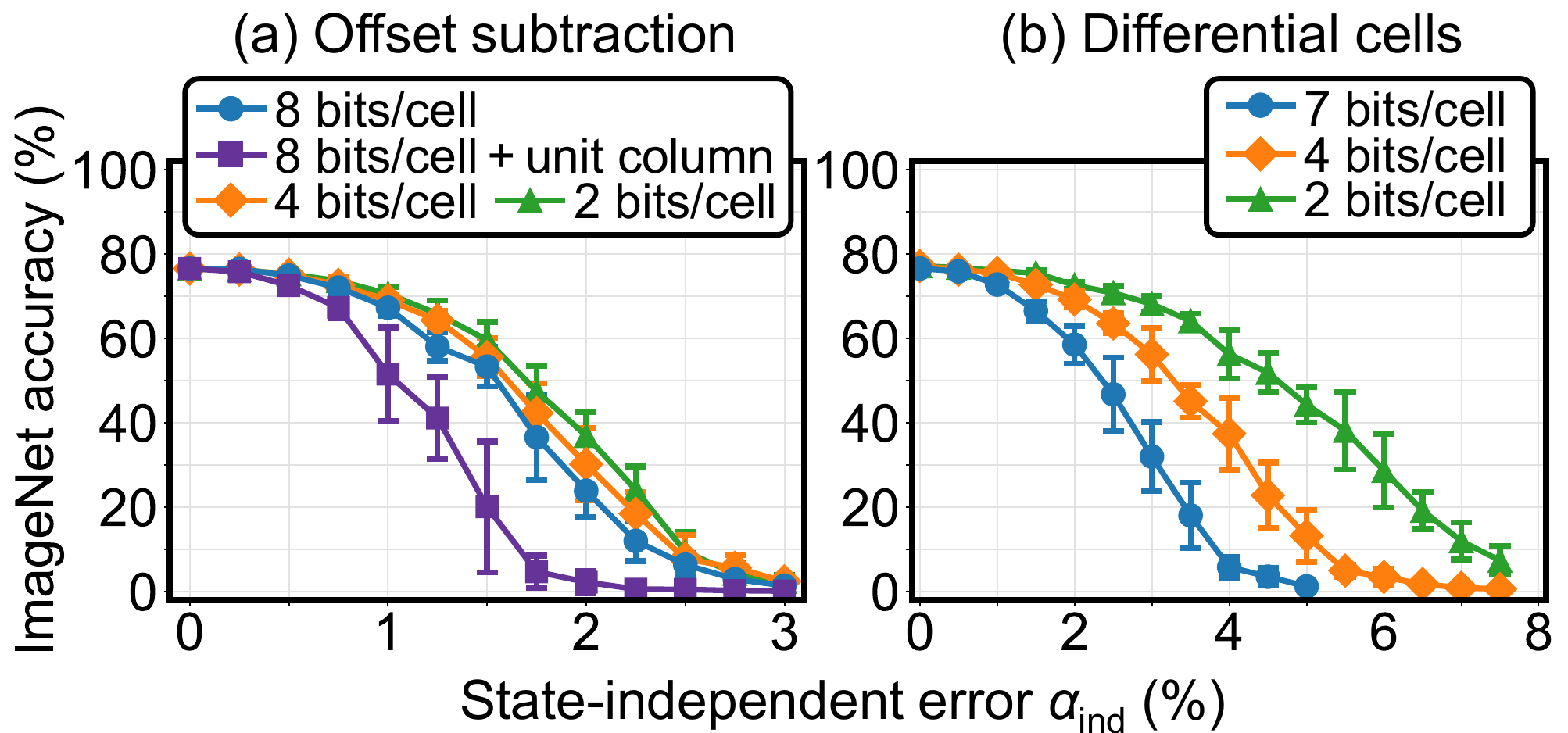}
    \caption{Sensitivity of ResNet50-v1.5 accuracy to state-independent errors using (a) arrays with offset subtraction and (b) arrays with differential cells. Error bars span two standard deviations over ten trials.
}
    \label{fig:independent_error}
   \vspace{-8pt}
\end{figure}

Fig.~\ref{fig:independent_error}(a) shows the accuracy sensitivity of offset-subtraction systems to state-independent errors, shown for different slice widths. In all cases, the accuracy is highly sensitive to error, falling nearly to zero at $\alpha_\text{ind} = 2.5\%$. The offset term to be subtracted is computed digitally, except in one case where a unit column is used.

The unit column is an additional column in the array whose conductances are all mapped to the center of the weight range\cite{Shafiee16}. The analog sum in this column is subtracted from all other sums. Fig.~\ref{fig:independent_error}(a) shows that this method incurs a large accuracy loss. The unit column accumulates error just as the other columns do, and this adds to the error in all other dot products when the offset is subtracted. By correlating the errors in these dot products, the unit column also increases the variance in the accuracy.

Fig.~\ref{fig:independent_error}(a) also shows that bit slicing can slightly improve accuracy. This is because the random programming errors in different bit slices can cancel. The amount of cancellation is limited, however, because the bit slices are not weighted equally when they are aggregated. The benefit of bit slicing can be analyzed in terms of the signal-to-noise ratio (SNR) of the dot product, following Genov and Cauwenberghs~\cite{Genov00}. 

As an example, consider offset subtraction with two bits per cell. The dot products in different slices add as in Equation \eqref{eq:bitslicing_offset}, while the errors in different slices add in quadrature. As a shorthand, let $D_i=\mat{W}_i\vec{x}$ denote the slice-wise dot products and $\sigma_i$ the errors in $D_i$ due to accumulated cell errors. The SNR prior to the digital offset subtraction is:
\vspace{-1pt}
\begin{align}
\label{eq:offset_error}
\text{SNR} = \frac{2^6 D_3 + 2^4 D_2 + 2^2 D_1 + D_0}{\sqrt{(2^6 \sigma_3)^2 + \left(2^4 \sigma_2 \right)^2 + \left(2^2\sigma_1\right)^2 + \sigma_0^2}}
\end{align}
This expression can be simplified. With offset subtraction, the expected values of $D_i$ are similar since every slice has close to the same average conductance (see Fig. \ref{fig:conductance_stats}). With state-independent errors, the expected errors $\sigma_i$ in each slice must also be the same. Therefore:
\vspace{-1pt}
\begin{align}
\label{eq:offset_error2}
\frac{\text{SNR}}{\text{SNR}_0} \approx \frac{2^6 + 2^4 + 2^2 + 1}{\sqrt{2^{12} + 2^8 + 2^4 + 1}} = 1.286
\end{align}
where $\text{SNR}_0 = D_0/\sigma_0$ is the SNR with unsliced weights. Using four slices slightly increases the dot product SNR. One can show that there is a theoretical maximum SNR increase of $\sqrt{3}$ relative to unsliced weights, obtained in the limit of infinitely many 1-bit slices. Bit slicing thus provides a small accuracy benefit, consistent with Fig. \ref{fig:independent_error}(a).

Genov and Cauwenberghs~\cite{Genov00} derived a further SNR benefit of up to $\sqrt{3}$ from \emph{input} bit slicing, assuming that dot product errors for the different input bits are also independent. In general, this is not true of programming errors, which are static between analog MVM operations. However, this is true of conductance errors caused by cycle-to-cycle read noise. In systems that use input bit slicing, read noise has a weaker effect than programming errors because noise in different input bits within the same weight slice can cancel.

Fig.~\ref{fig:independent_error}(b) shows that differential cells are more tolerant to state-independent errors. This is because by using two cells per slice, the dot product SNR improves: the signal range doubles, but the dot product error increases only by $\sqrt{2}$. Also, the improvement in accuracy with the number of bit slices is larger than for the offset case. Recall from Fig. \ref{fig:conductance_stats} that unlike offset subtraction, the average conductance falls in value from the lowest to highest slice when using differential cells. This trend counteracts the exponential weighting of bit slices in Equation \eqref{eq:offset_error} so that the dot products in different slices are more equal in value. This enables greater cancellation of the dot product errors to occur. Because the assumptions that lead to Equation \eqref{eq:offset_error2} do not fully hold, the SNR improvement from bit slicing can exceed $\sqrt{3}$.

\subsection{Sensitivity to State-Proportional Errors}
\label{subsec:proportional_error}

\begin{figure}[t]
    \centering
    \includegraphics[width=0.48\textwidth]{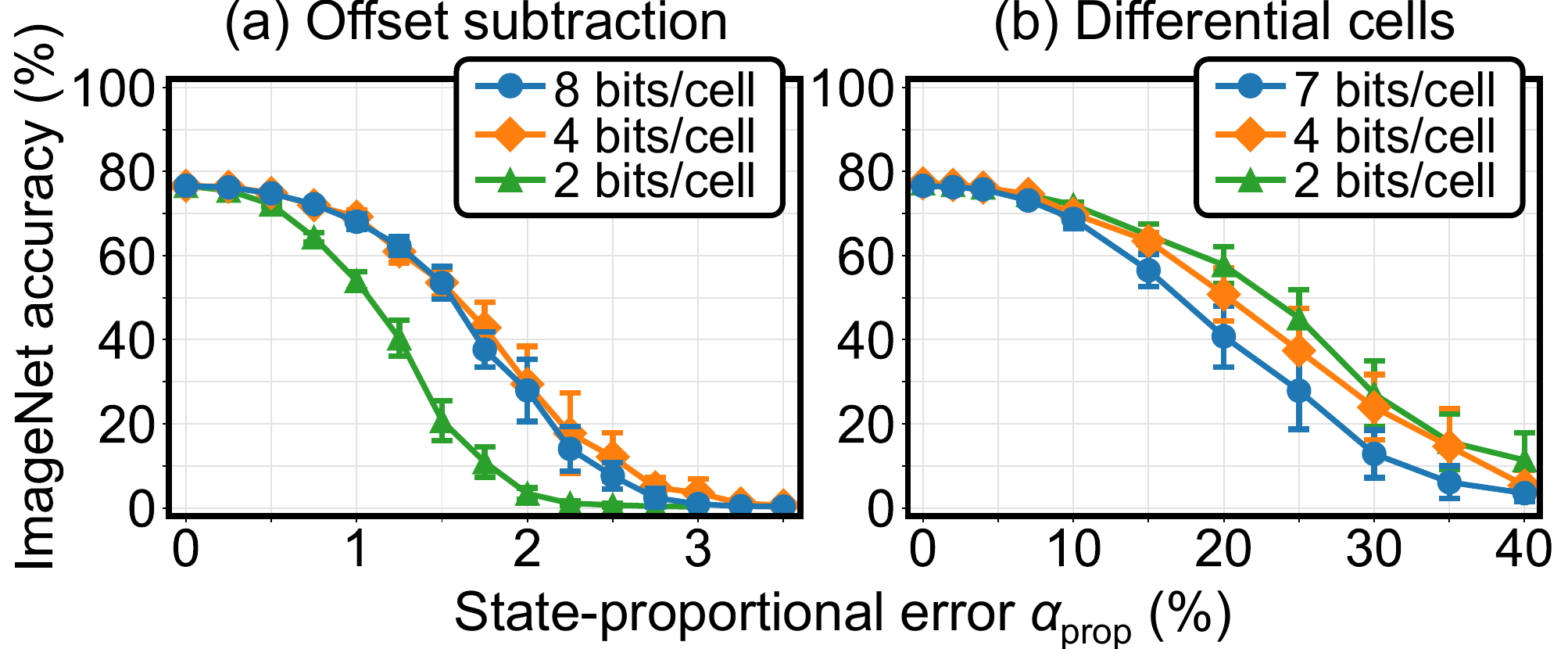}
    \caption{Sensitivity of ResNet50-v1.5 accuracy to state-proportional errors using (a) arrays with offset subtraction and (b) arrays with differential cells.
}
    \label{fig:proportional_error}
\end{figure}

Fig.~\ref{fig:proportional_error}(a) shows the sensitivity of offset subtraction systems to state-proportional cell errors. These systems lack proportionality between weights and conductances, and thus do not substantially discriminate between state-independent and state-proportional errors. A comparison of Fig.~\ref{fig:independent_error}(a) and Fig.~\ref{fig:proportional_error}(a) reveals that the sensitivity is similar between the two error types, but in Fig.~\ref{fig:proportional_error}(a), better accuracy is obtained with fewer slices. This is a subtle effect that stems from the bit representation used by the offset scheme in Fig. \ref{fig:array_setup}(c). Specifically, small negative weights use a conductance of $G_\text{max}$ in the second most significant slice, so that dot products that use more negative weights will have high state-proportional errors. This slice is weighted more heavily in systems with more slices, leading to a lower accuracy.

Fig.~\ref{fig:proportional_error}(b) shows that \emph{a system with differential cells is very tolerant to state-proportional errors, with $>$10$\times$ the resilience of offset-subtraction systems.} This large difference results from proportionality; as shown in Fig.~\ref{fig:weight_dists}, most of the weights in ResNet50-v1.5 are close to zero. Consequently, Fig.~\ref{fig:conductance_stats} shows that for both unsliced weights and the top slice in bit-sliced systems, the average cell conductance is a small fraction of $G_\text{max}$ and thus has a small error. The lower slices have larger errors, but these are suppressed by the S\&A operation. As arrays with fewer bits per cell are used, the top slice becomes more zero-dominated, reducing dot product errors and enabling higher accuracy. Importantly, for bit-sliced systems, a large fraction of the improvement from Fig.~\ref{fig:independent_error}(b) to Fig.~\ref{fig:proportional_error}(b) can be attributed to the error reduction specifically in the minimum conductance state $G_\text{min}$.

\begin{figure}[t]
    \centering
    \includegraphics[width=0.48\textwidth]{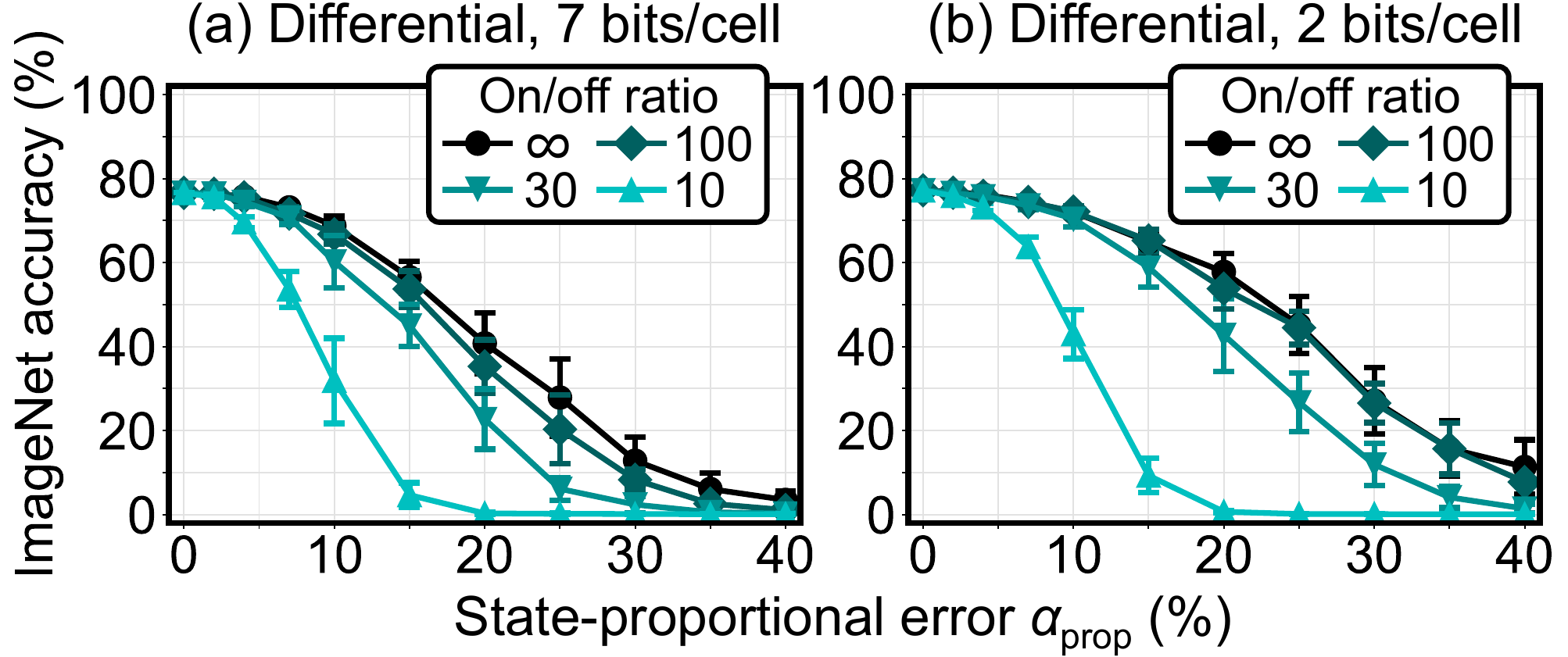}
    \caption{Sensitivity of ResNet50-v1.5 accuracy to state-proportional errors and cell On/Off ratio, using differential cells (a) without bit slicing and (b) with bit slicing.}
    \label{fig:on_off}
   \vspace{-8pt}
\end{figure}

When the cell has a finite On/Off ratio ($G_\text{min} > 0$), there is only partial proportionality between the weight magnitudes and cell conductances, and current can flow through cells that encode zero-valued weights. Fig.~\ref{fig:on_off} shows that a low On/Off ratio increases the sensitivity to state-proportional errors, but an On/Off ratio of 100 has nearly the same resilience as an infinite On/Off ratio. The effect is similar in differential cells with and without bit slicing.

With an On/Off ratio of 100 or more, systems with differential cells see nearly zero accuracy loss for state-proportional errors below $\alpha_\text{prop}=5\%$, even if these errors are allowed to accumulate in an array as large as the weight matrix (up to 4608 rows). The high-accuracy regions of the sensitivity curves in Fig.~\ref{fig:independent_error} to \ref{fig:on_off} correspond to the regime where direct weight transfer can be used with negligible accuracy penalty. Retraining is expected to be useful when the cell error falls in the intermediate-accuracy regions of these curves.

\subsection{Error Sensitivity vs. Neural Network}
\label{subsec:networks}

\begin{figure}[t]
    \centering
    \includegraphics[width=0.48\textwidth]{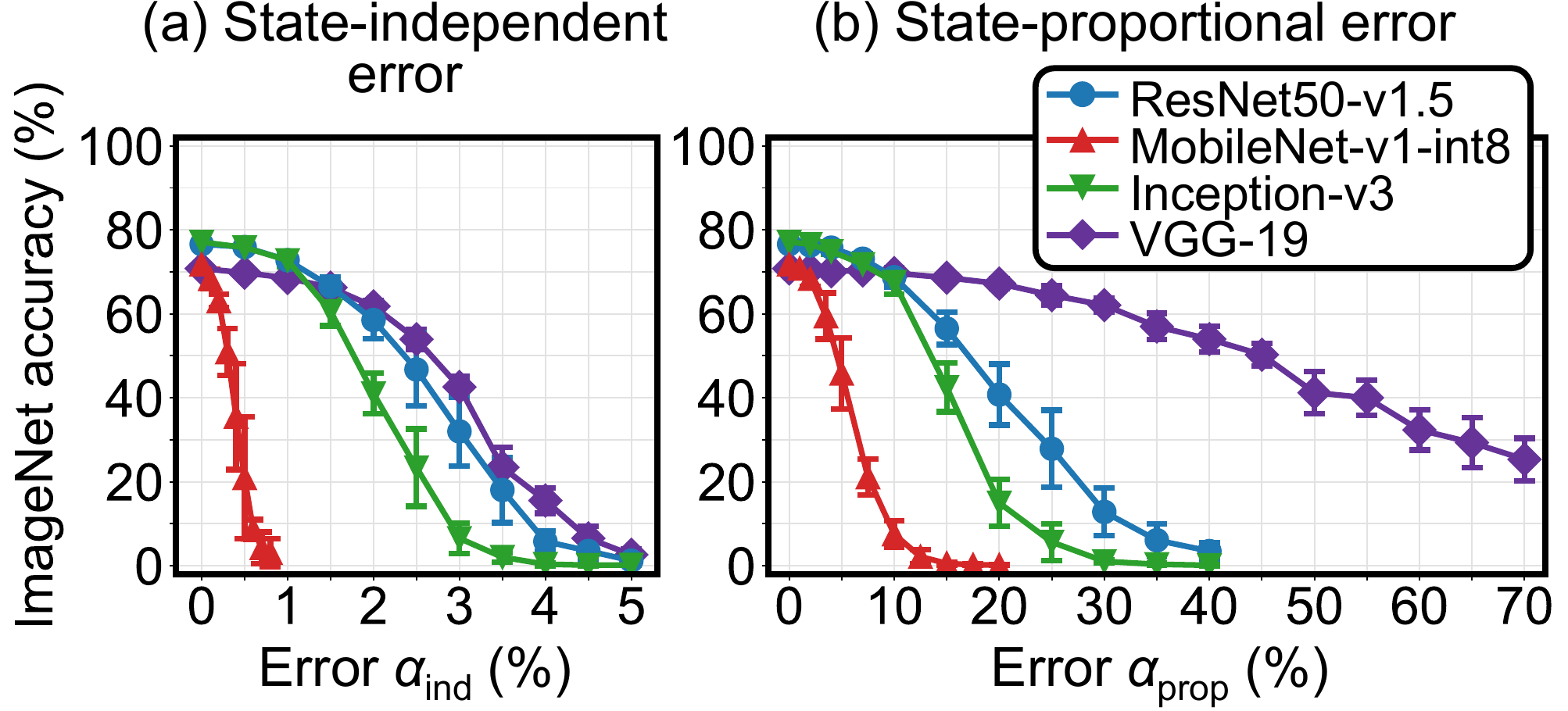}
    \caption{Sensitivity of four ImageNet neural networks to (a) state-independent errors and (b) state-proportional errors. The results assume differential cells without bit slicing and infinite On/Off ratio.}
    \label{fig:networks}
   \vspace{-8pt}
\end{figure}

Fig.~\ref{fig:networks} generalizes the conclusions from the previous sections to three other ImageNet neural networks, whose weights are all quantized to 8 bits. All of the evaluated networks have a much weaker sensitivity to state-proportional errors than state-independent errors. This results from the proportionality of both the cell conductance and the conductance error to the weight value, combined with the fact that all of the networks have a strongly zero-peaked weight value distribution as shown in Fig.~\ref{fig:weight_dists}. To obtain both types of proportionality, analog accelerators should use cell technologies with both \emph{state-proportional errors and high On/Off ratio}. Field-effect transistor memories such as flash can fulfill both requirements, as we show in Section \ref{sec:sonos}.

The differences in sensitivity across the four networks can be explained to first order based on the number of parameters, listed in Table \ref{tab:networks}. MobileNet has by far the fewest weights and thus the least amount of redundancy in its information content; therefore, its accuracy is more sensitive to errors in these weights. The opposite is true for VGG-19, which has the most weights and is thus the most error-tolerant\cite{Zhang18}. The strong sensitivity of VGG-19 to state-independent errors, relative to its larger model size, is due to a very large fully-connected layer (25088 matrix rows) with a large amount of error accumulation. With state-proportional errors, the large matrix size is less consequential, since most of the elements have small or zero values.

Fig.~\ref{fig:networks} demonstrates the intuitive result that error tolerance can be achieved at the algorithm level by using a larger network with more redundant parameters, such as VGG-19. However, a larger network requires more energy and area to deploy. Reducing the size of the state-proportional cell error $\alpha_\text{prop}$ allows similar (or superior) accuracy to be achieved using a network with a smaller footprint.

\section{Robustness to Quantization Errors}
\label{sec:ADC}

This section shows that by digitizing analog outputs at a precision that matches the network's inherent precision requirements, the ADC requirements can be relaxed to the minimum feasible resolution that still yields high accuracy. The analog output is assumed here to be a voltage $V$.

\subsection{ADC Errors}
\label{subsec:ADC_errors}

\begin{figure}[t]
    \centering
    \includegraphics[width=0.5\textwidth]{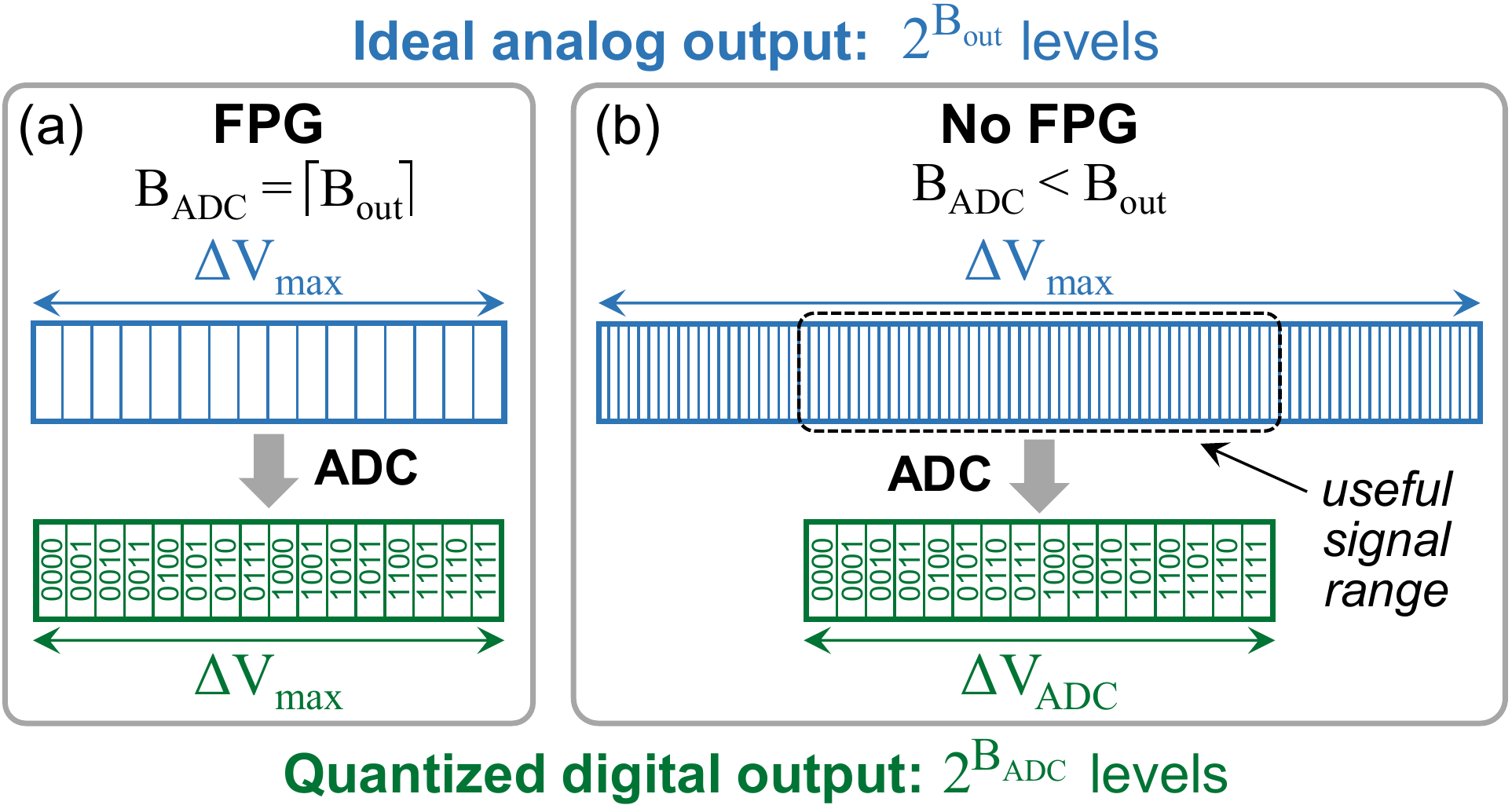}
    \caption{(a) The FPG provides a one-to-one mapping between possible analog outputs and ADC levels. (b) When not using the FPG, the range and resolution of the ADC can be chosen independently from those of the analog signal to minimize quantization and clipping errors.
}
    \label{fig:adc_errors}
   \vspace{-8pt}
\end{figure}

The bit resolution $B_\text{out}$ given by Equation \eqref{eq:full_precision} is the resolution contained in the analog output of the array if the analog computation were free of errors. Equivalently, $B_\text{out}$ measures the amount of computation that is done by the array before leaving the analog domain. Under the FPG, the ADC resolution is set equal to $B_\text{out}$, so that there is a one-to-one mapping from the possible analog outputs to the ADC's digital levels, as shown in Fig. \ref{fig:adc_errors}(a). Ideally, this guarantees no loss of information upon digitization. However, because the ADC resolution must be kept moderately low (typically 8 bits) due to energy considerations, the FPG limits the amount of analog processing and its associated energy benefits.

Alternatively, the ADC resolution $B_\text{ADC}$ can be kept well below $B_\text{out}$. In this case, the ADC \emph{compresses} a higher-resolution analog output into a lower-resolution digital output, as shown in Fig. \ref{fig:adc_errors}(b). This compression induces two potential kinds of error: (1) quantization error, due to the potentially larger separation between ADC levels compared to the minimum separation between analog output levels, and (2) clipping error, if the signal range spanned by the ADC is smaller than the possible range of the signal. Any signal lying outside the ADC range is assumed to clip to the highest or lowest ADC level.

\subsection{Calibrating the ADC Range}
\label{subsec:ADC_distribution}

A way to eliminate clipping errors altogether is to make the ADC range equal to the maximum possible range of output voltages: $\Delta V_\text{ADC} = \Delta V_\text{max}$. In a practical inference application, however, the analog outputs may be much smaller on average than $\Delta V_\text{max}$, and this design choice would leave most of the ADC levels heavily underutilized, increasing quantization errors. To better utilize the ADC levels, the range of values quantized by the ADC should be calibrated to the \emph{useful} range of the analog signal, as illustrated in Fig. \ref{fig:adc_errors}(b).

\begin{figure}[t]
    \centering
    \includegraphics[width=0.4\textwidth]{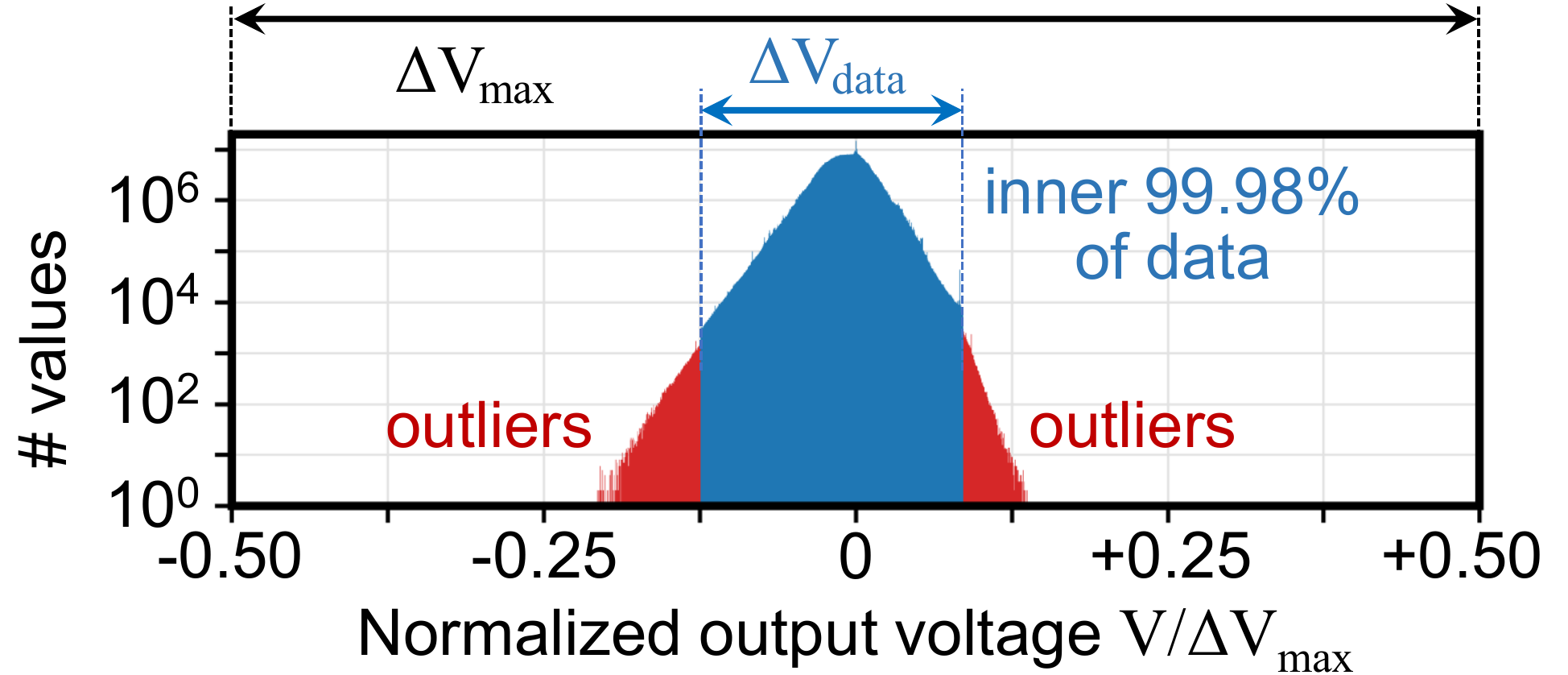}
    \caption{Distribution of output voltages for the sixth convolution layer of ResNet50-v1.5 (res2b\_branch2b), using differential cells with unsliced weights, collected using the MLPerf calibration set.
}
    \label{fig:adc_inputs}
   \vspace{-3pt}
\end{figure}

\begin{figure}[t]
    \centering
    \includegraphics[width=0.38\textwidth]{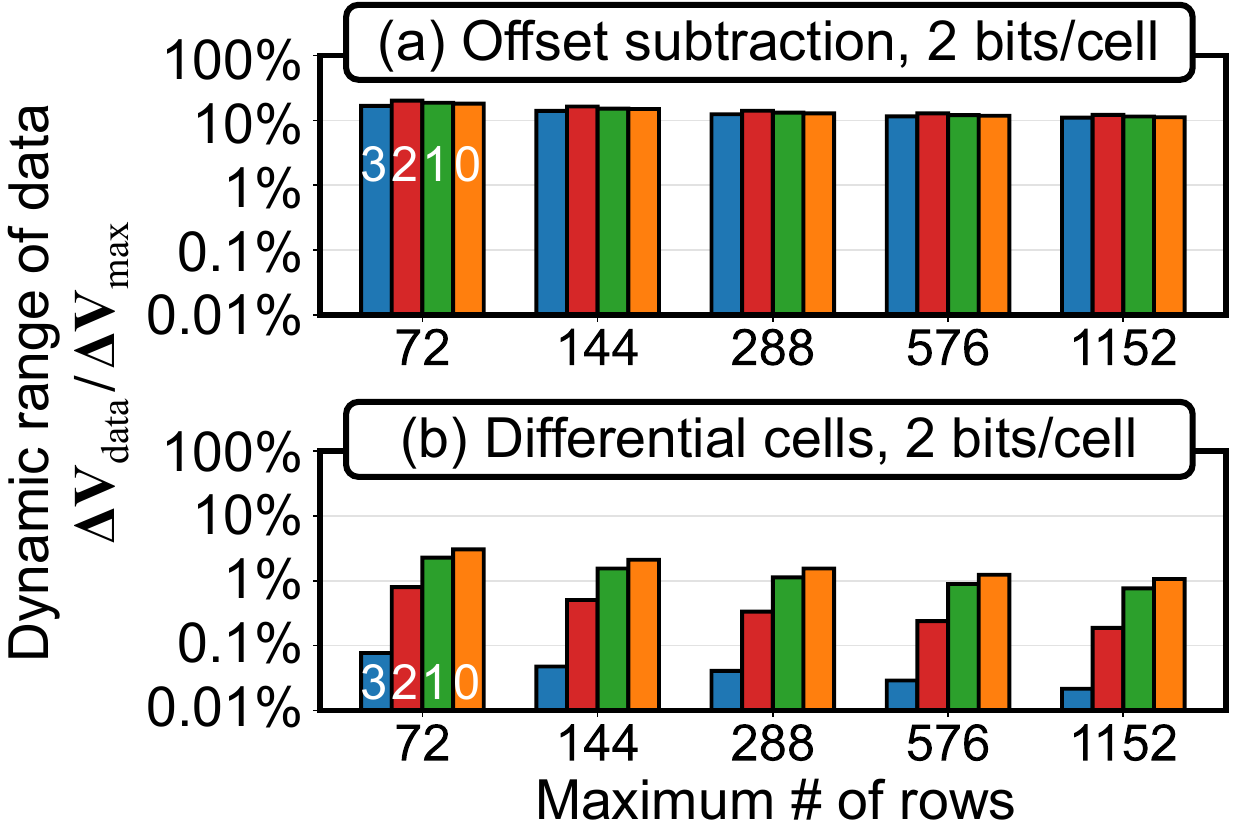}
    \caption{Output voltage statistics for ResNet50-v1.5 with (a) offset subtraction and (b) differential cells, using 2 bits/cell, with different constraints on array size. Ranges are averaged over all layers. Bars are colored by bit slice index (0 = lowest).
}
    \label{fig:adc_clip}
   \vspace{-9pt}
\end{figure}

Our process for determining the optimal ADC quantization ranges is qualitatively similar to that used by Jacob \etal{} for low-precision digital inference\cite{Jacob18}, where the optimal activation quantization ranges are found from the activation statistics seen during training. Similar ideas have also been proposed for analog systems\cite{Joshi20,Gonungondla20}. We collect the statistics on every array's output voltages (ADC inputs) by simulating inference on the MLPerf calibration subset of 500 ImageNet images\cite{MLPerfCalibrationGit}. The ADC limits of each layer are separately calibrated, as are the ADC limits of different bit slices within a layer, whose outputs can differ greatly in range.

Fig.~\ref{fig:adc_inputs} shows an example distribution of normalized output voltages, for a layer in ResNet50-v1.5. ADC range calibration relies on a single statistical property of these distributions: the range $\Delta V_\text{data}$ that contains the inner $P=99.98\%$ of all collected values of $V$. This was empirically determined to be the useful signal range for ResNet50-v1.5, as clipping the remaining 0.02\% of outlier values had a negligible effect on accuracy. The ADC limits are chosen to be just large enough to contain the useful signal range (i.e. $\Delta V_\text{ADC} \geq \Delta V_\text{data}$). For bit-sliced systems, the ADC limits of different slices are constrained to differ only by a power of two; this ensures that their results can still be aggregated via S\&A operations without any complex scaling steps. With unsliced weights, there is no such constraint on the ADC limits. We note that in general, the definition of the useful signal range (set by the single parameter $P$) may need to be tuned to optimize the accuracy for a given neural network, dataset, mapping scheme, and ADC resolution.

Comparing the useful signal range $\Delta V_\text{data}$ (normalized to $\Delta V_\text{max}$) of different mapping schemes and bit slices reveals important insights about their output voltage distributions: this is shown in Fig.~\ref{fig:adc_clip}. When using offset subtraction, as in Fig.~\ref{fig:adc_clip}(a), the useful signal occupies 10-20\% of $\Delta V_\text{max}$; the remainder is used only by outlier values. This somewhat low percentage results from the fact that input activations tend to concentrate near zero, especially in ReLU networks\cite{Miyashita16}. Thus, the ADC levels can be safely re-allocated to cover only this smaller range to offer better signal resolution.

Fig.~\ref{fig:adc_clip}(b) shows that the useful signal range is orders ofmagnitude smaller for differential cells: less than 0.1\% of $\Delta V_\text{max}$ for the top slice. This is principablly a result of proportional mapping: since differential cells use much lower conductances as shown in Fig.~\ref{fig:conductance_stats}, the output voltages are reduced correspondingly. There is also a significant signal reduction from the analog cancellation of positive and negative bit line currents. The smaller signal range enables a much more aggressive reduction of the ADC range. 

\subsection{Matching the ADC to the Algorithm's Precision}
\label{subsec:ADC_precision}

\begin{figure}[t]
    \centering
    \includegraphics[width=0.495\textwidth]{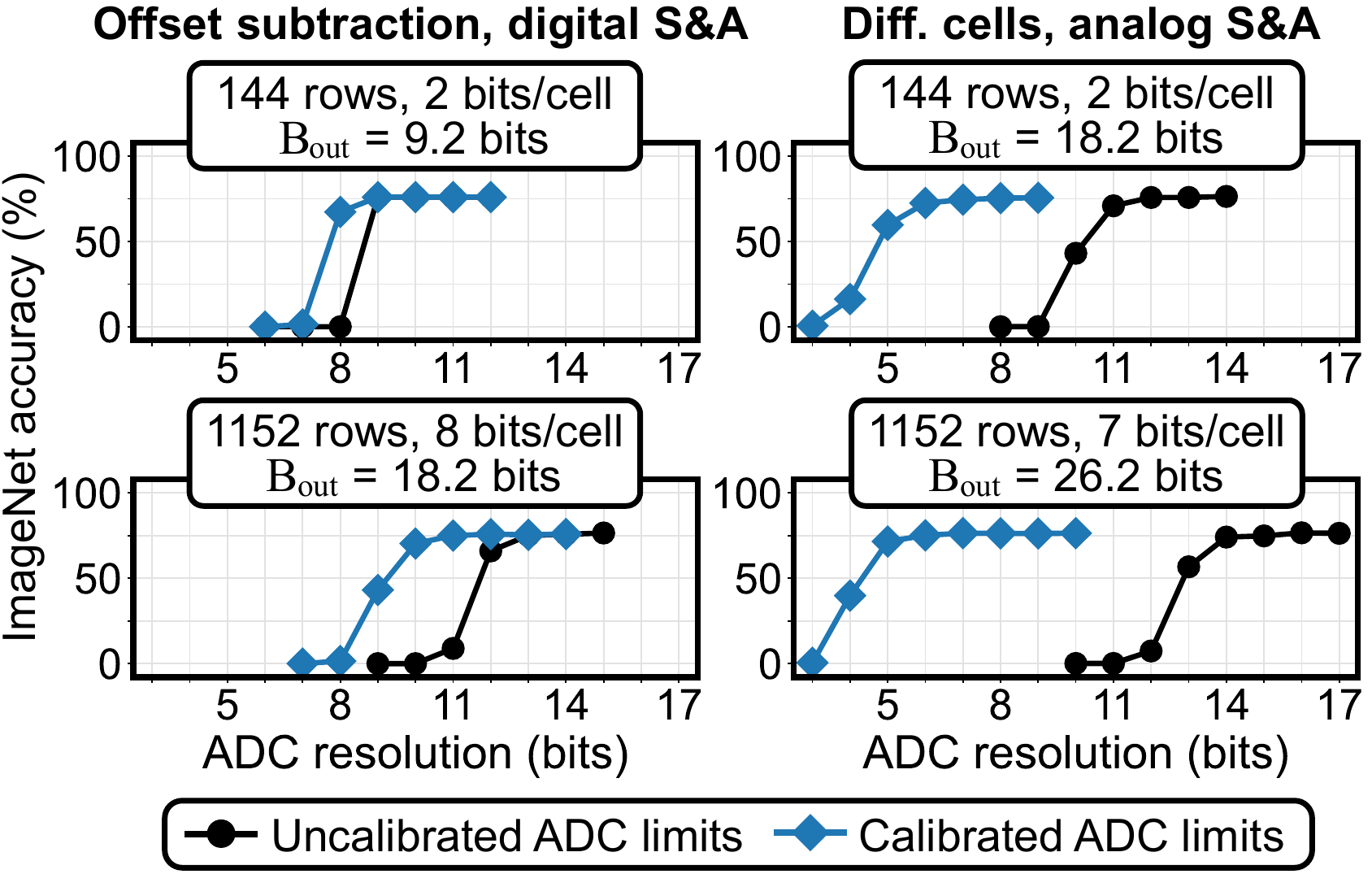}
    \caption{ImageNet accuracy using ResNet50-v1.5 vs ADC resolution for different weight mapping schemes, without calibrated ranges ($\Delta V_\text{ADC} = \Delta V_\text{max}$) and with calibrated ranges ($\Delta V_\text{ADC} \approx \Delta V_\text{data}$).
}
    \label{fig:adc_accuracy}
    \vspace{-8pt}
\end{figure}

Fig.~\ref{fig:adc_accuracy} shows the ADC resolution sensitivity of ImageNet accuracy for different mapping schemes. ADC quantization is assumed to be deterministic, and cell errors are not included in order to isolate the ADC's effect. As described in Section \ref{subsec:input_bits}, input bits are aggregated with digital circuitry for offset subtraction ($B_\text{in} = 1$ bit) and with analog circuitry for differential cells ($B_\text{in} = 8$ bits). This  leads to a much higher analog resolution $B_\text{out}$ for differential cells.

In all cases, calibration of the ADC range allows high accuracy to be obtained at a reduced resolution. By confining the ADC range to the useful signal range, quantization errors are reduced for the same number of levels. The benefit of calibration is greater for differential cells, which have a much smaller useful signal range as explained in Section \ref{subsec:ADC_distribution}.

Fig.~\ref{fig:adc_accuracy} further shows that after range calibration is applied, differential cells can tolerate an ADC with several fewer bits of resolution than offset subtraction systems: evidently, differential cells are more resilient to quantization errors. This resilience can be understood by again considering the proportional weight mapping property of differential cells. Combined with the analog subtraction of currents, this results in \emph{dot product proportionality}: the voltages at the ADC input are proportional to the numerical values of the dot products (or the slice-wise dot products). Offset subtraction systems lack this critical proportionality, since an offset must be subtracted after the ADC to obtain the true dot products.

Dot product proportionality implies that the data compression function of the ADC is effectively applied to the numerical dot products. Therefore, the required ADC resolution is directly connected to \emph{the neural network's inherent sensitivity to data precision}, which is hardware-independent and is fully decoupled from $B_\text{out}$. The effect is most striking with unsliced weights, where $B_\text{out} = 26.2$ bits but high accuracy is maintained down to $B_\text{ADC} = 7$ bits. This is close to the inherent precision sensitivity of ImageNet neural networks, which is typically about 8 bits\cite{Jacob18}.

\begin{figure}[t]
    \centering
    \includegraphics[width=0.45\textwidth]{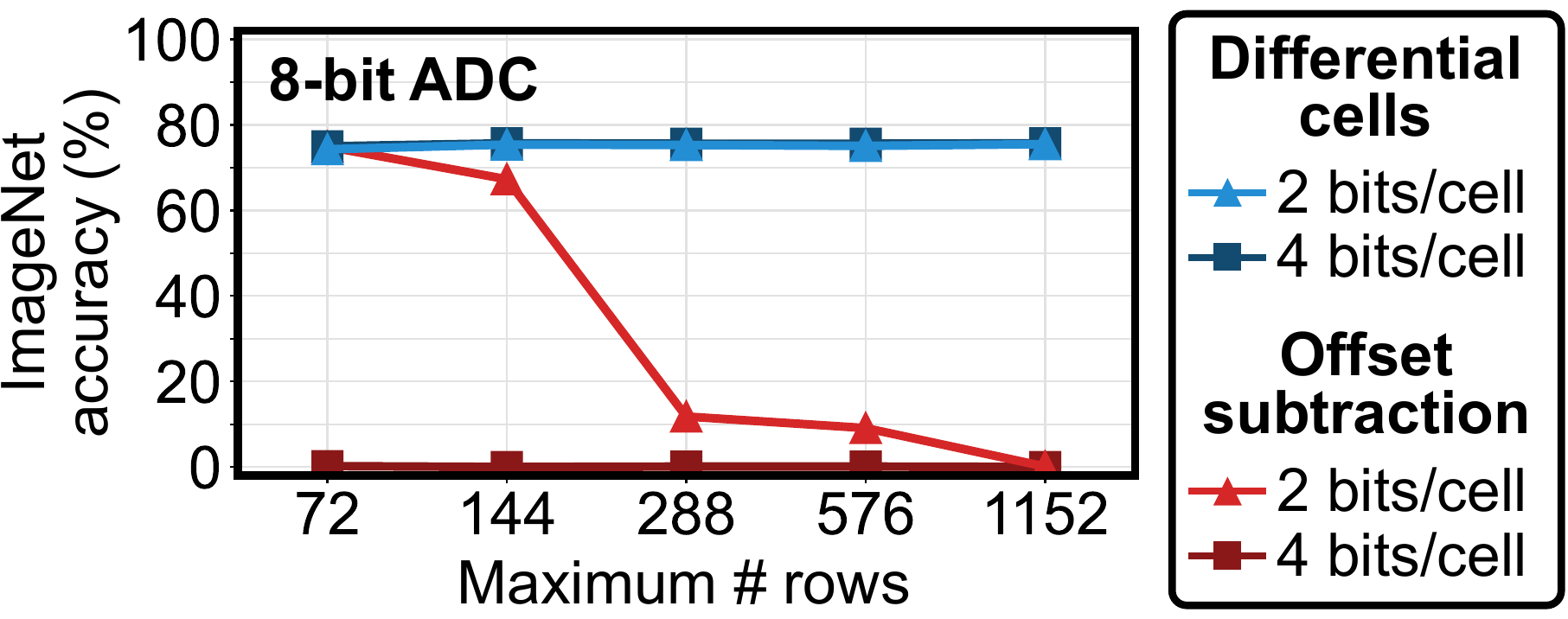}
    \caption{ImageNet accuracy using ResNet50-v1.5 with a calibrated 8-bit ADC for different mapping schemes, cell precision, and array size. The ADC range was separately calibrated for each point.
}
    \label{fig:size_accuracy}
   \vspace{-8pt}
\end{figure}

Fig.~\ref{fig:size_accuracy} compares systems with offset subtraction (without dot product proportionality) and differential cells (with dot product proportionality) at a fixed ADC resolution of 8 bits. Offset-subtraction systems can only tolerate an 8-bit ADC when the array is small ($\leq$144 rows) and the weights are finely sliced ($\leq$2 bits/cell), which together bring $B_\text{out}$ close to 8 bits. Differential cells suffer almost no accuracy loss with an 8-bit ADC regardless of the bits per cell and array size; this again illustrates that the accuracy is decoupled from $B_\text{out}$. Dot product proportionality makes a practical ADC resolution of 8 bits compatible with a much larger analog resolution $B_\text{out}$. Equivalently, much more computation can be done in the analog domain before the signal is ever converted to digital. This has significant consequences for energy efficiency, discussed in Section \ref{sec:sonos}.

\section{Suppressing Error Propagation}
\label{sec:quant}

\begin{figure}[t]
    \centering
    \includegraphics[width=0.41\textwidth]{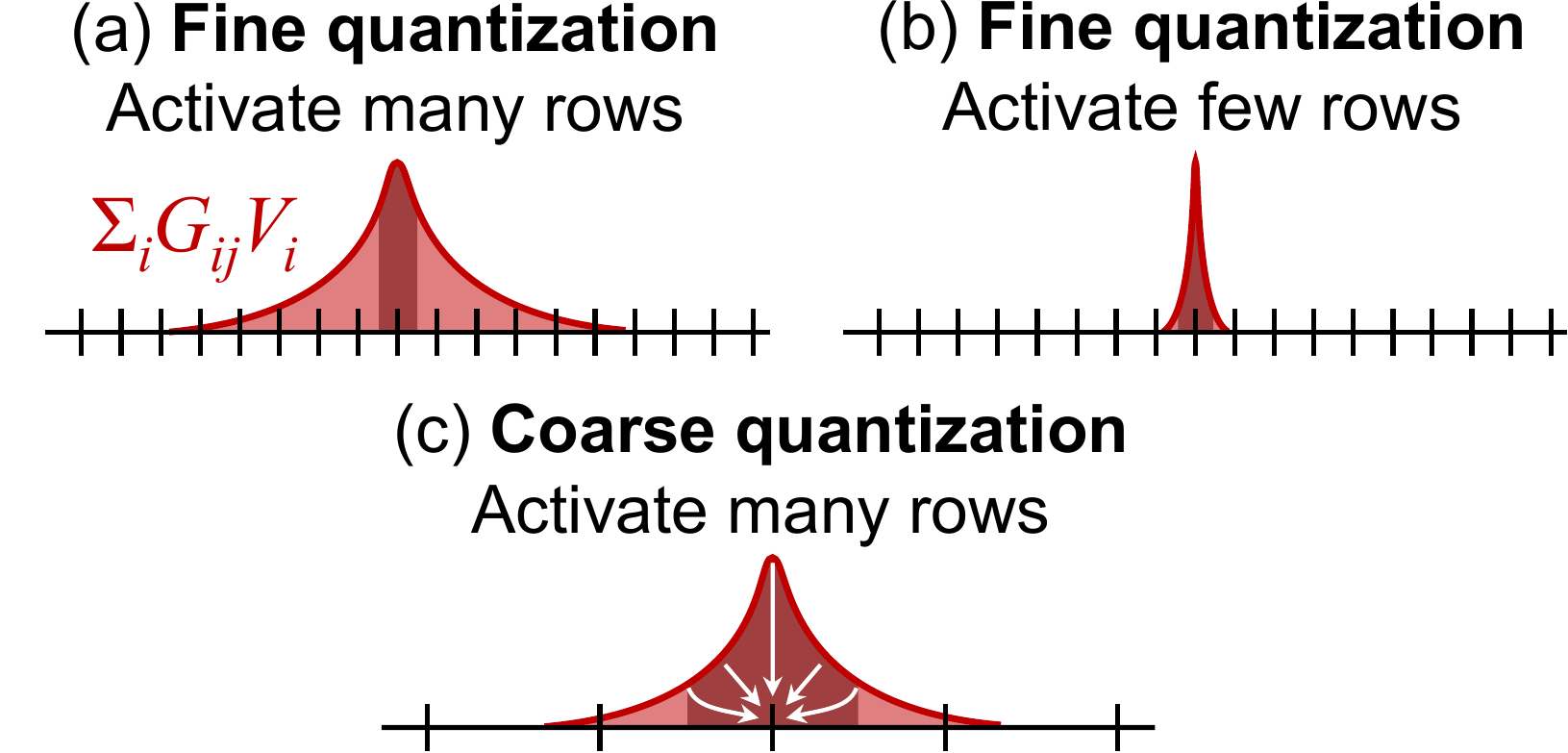}
    \caption{(a) The distribution of accumulated cell errors can span multiple digital levels. Propagation of these errors can be suppressed if they fall within a single level. For the same cell error, this can be achieved by (b) activating fewer rows per MVM or by (c) using coarse ADC or activation quantization.
}
    \label{fig:coarse_quant}
\end{figure}

\begin{figure}[t]
    \centering
    \includegraphics[width=0.48\textwidth]{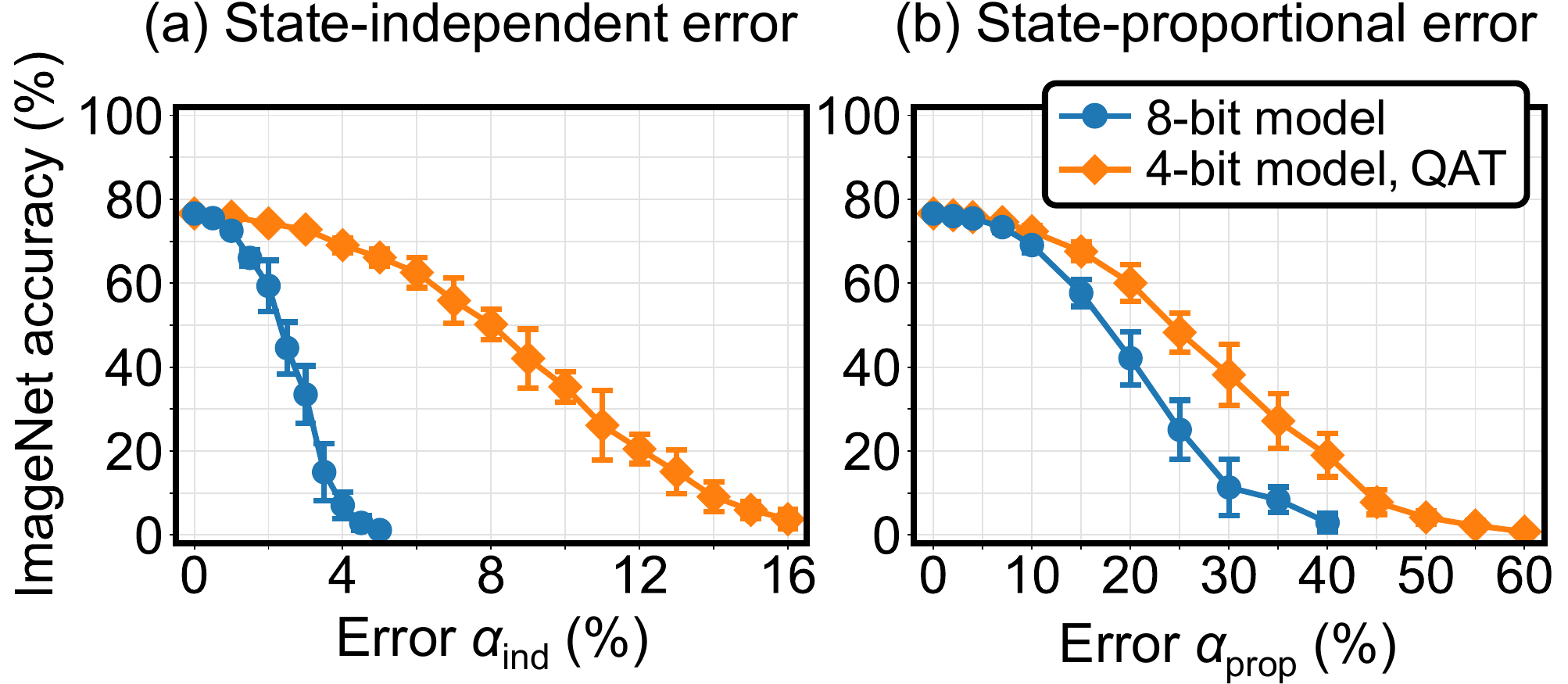}
    \caption{Sensitivity of two ResNet50-v1.5 models at 8-bit and 4-bit precision to (a) state-independent errors and (b) state-proportional errors. Simulations include calibrated 8-bit ADCs and assume differential cells with unsliced weights and 1152 maximum rows.
}
    \label{fig:int4_error}
   \vspace{-8pt}
\end{figure}

As shown in Section~\ref{sec:errors}, the accuracy loss due to accumulated cell errors can be minmized by using sufficiently precise memory cells and exploiting state-proportional errors. With less precise cells, some prior work has relied on ADC quantization to cut off the propagation of cell errors from layer to layer in a DNN\cite{Hu16,Yang19,Zhang20}. Yang \etal{}~\cite{Yang19} activated only a few rows per MVM, such that on average, the accumulated errors on a bit line fall below the separation of levels in an ADC, as shown in Fig. \ref{fig:coarse_quant}(b). While this approach succeeds in suppressing error propagation, it reduces energy efficiency since many more analog MVMs (and ADC operations) are needed to process each layer. A coarse ADC can provide the same benefit without reducing the number of rows, as shown in Fig. \ref{fig:coarse_quant}(c), but the accuracy would suffer due to quantization errors, as discussed in Section~\ref{sec:ADC}. A purely hardware solution cannot solve this problem, but it is possible to eliminate the quantization errors by training a DNN to tolerate \emph{low-precision activations} during inference. This would combine the benefits of low quantization errors, greater resilience to cell errors, and high energy efficiency.

Unlike training techniques that are specialized for analog systems (see Section \ref{sec:background-retraining}), quantization-aware training (QAT) benefits digital accelerators by reducing the computational load at inference time. Therefore, there has been much recent work on 4-bit or lower resolution networks with nearly no accuracy loss relative to floating-point networks~\cite{Choi18,Zhang18,Sun20}. Importantly, the broad applicability of low-precision networks increases the likelihood that QAT methods can be integrated into state-of-the-art training workflows.

This section evaluates a 4-bit QAT network with the ResNet50-v1.5 topology, submitted by Nvidia to the MLPerf Inference Benchmark~\cite{NvidiaInt4}. The network uses 4-bit weights and activations in all layers except the first and last, which use 8-bit weights. Each ReLU output is multiplied by 16-bit scaling factors before quantizing to 4 bits. The digital software accuracy of this network is shown in Table \ref{tab:networks}. When simulating the analog accuracy of this network, these scaling steps are processed digitally between \insitu{} MVMs.

Fig. \ref{fig:int4_error} compares the error sensitivity of the 4-bit QAT model with the floating-point ResNet50-v1.5 model, whose weights are quantized to 8 bits after training. For a fair comparison, an 8-bit ADC is included for both cases; in the 4-bit model, this higher-resolution ADC helps minimize errors prior to the 4-bit quantization step, which is performed digitally. Fig. \ref{fig:int4_error}(a) shows that the 4-bit model is substantially more resilient to state-independent errors than the 8-bit network. This results entirely from activation quantization, and not weight quantization. The large separation between the 16 activation levels effectively cuts off the propagation of accumulated cell errors from one layer to the next. The same cell error results in a smaller dot product error on average. 

Fig. \ref{fig:int4_error}(b) shows that the 4-bit network is also more resilient to state-proportional error, but here the benefit is smaller. This can be explained by the different weight value distributions of the two networks. With only 16 levels, the distribution of the 4-bit weights cannot peak as sharply at zero as the 8-bit weights, which have 256 levels. As a result, the memory cells in the 4-bit network have a significantly higher average conductance (7.52\% of $G_\text{max}$) than the cells in the 8-bit network (1.95\% of $G_\text{max}$). Thus, deeply quantized \emph{weights} can actually be harmful for analog systems, as it leads to a higher state-proportional error per cell. This is ultimately outweighed by the benefit of quantized activations, so there is a net improvement in error sensitivity.

Notably, for the 8-bit network, the sensitivity to both types of error with an 8-bit ADC remains largely unchanged from the case with no ADC quantization, shown in Fig. \ref{fig:independent_error}(b) and Fig. \ref{fig:proportional_error}(b). As shown in Section \ref{sec:ADC}, a calibrated 8-bit ADC on its own induces very little accuracy loss for this network. When the ADC levels are relatively finely spaced, as depicted in Fig. \ref{fig:coarse_quant}(a), deterministic quantization neither suppresses nor compounds the effect of accumulated cell errors.

\section{Mitigating Parasitic Resistance}
\label{sec:parasitics}

Errors in the current conducted by a cell can arise not only from conductance errors, but also from voltage errors. A major source of voltage errors is the parasitic metal resistance, which induces voltage drops along the array's rows and columns. The resulting errors in the cell currents are spatially non-uniform and input-dependent. The effect grows super-linearly with array size, as each new row contributes both a line resistance and a source of current. Together with accumulated cell errors, this effect limits the size of an \insitu{} MVM. It is well known that parasitic resistance degrades MVM accuracy and some compensation methods have been proposed~\cite{Hu16,Zhang20,Jeong18,Jain19}. Here, we evaluate the end-to-end accuracy impact of parasitic voltage drops and their dependence on architecture-level design choices.

\begin{figure}[t]
    \centering
    \includegraphics[width=0.44\textwidth]{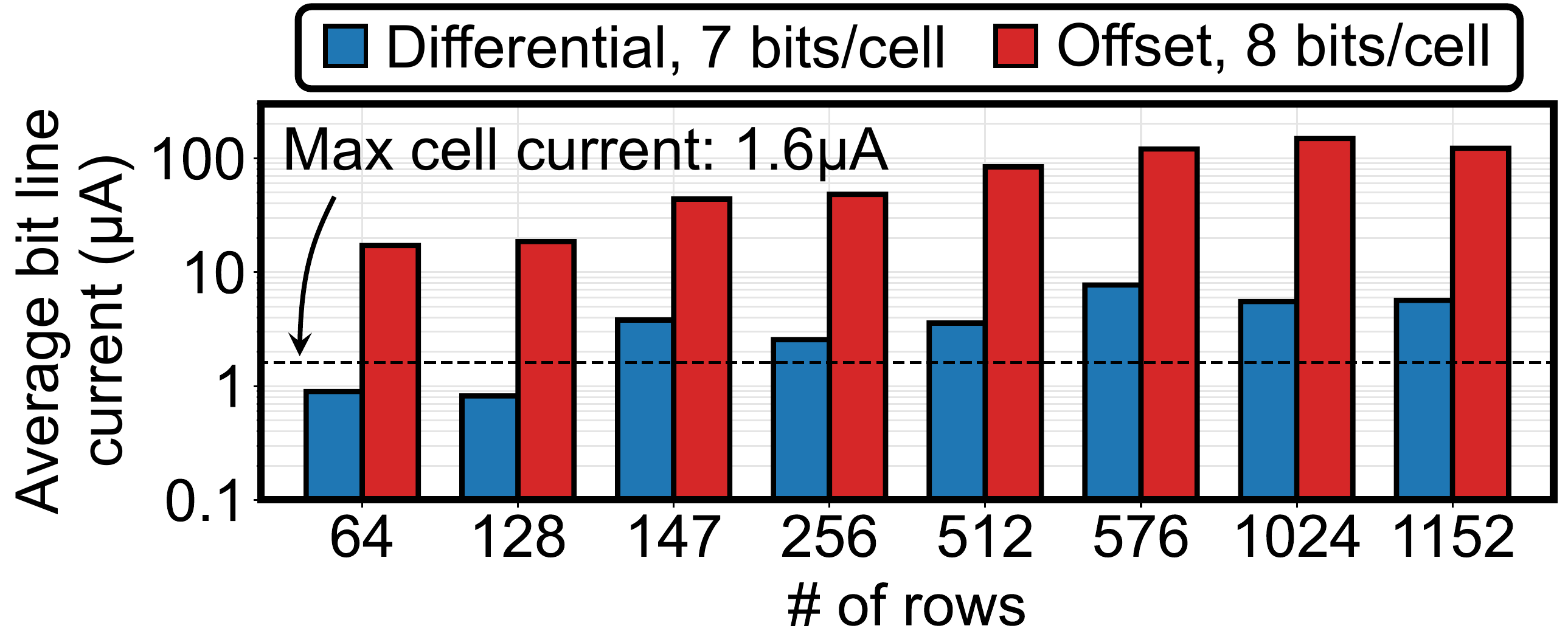}
    \caption{Bit line currents vs array size and mapping for ResNet50-v1.5. Layers that use the same \# rows are averaged together. Inputs are applied bit-wise based on the cell in Fig. \ref{fig:MVM}(c), and results are computed for the input LSB, which activates the most rows. The cell current varies from 0 to $I_\text{max}$ = 1.6 $\upmu$A (infinite On/Off ratio).}
    \label{fig:currents}
   \vspace{-8pt}
\end{figure}

Since parasitic voltage drops are proportional to bit line currents, they can be compared to state-proportional cell errors, described in Section~\ref{subsec:proportional_error}. Like state-proportional cell errors, parasitic resistance errors can be reduced by using a proportional mapping. Proportionality exploits the zero-peaked distribution of the weights to reduce the average cell conductance, and hence the accumulated bit line currents. Fig. \ref{fig:currents} shows that for ResNet50-v1.5, differential cells reduce the average bit line current by more than an order of magnitude. Even in arrays with as many as 1152 rows, the average bit line current (before analog subtraction) is only a few times the maximum current of a single cell. The lower bit line currents directly lead to smaller parasitic voltage drops.

\begin{figure}[t]
    \centering
    \includegraphics[width=0.46\textwidth]{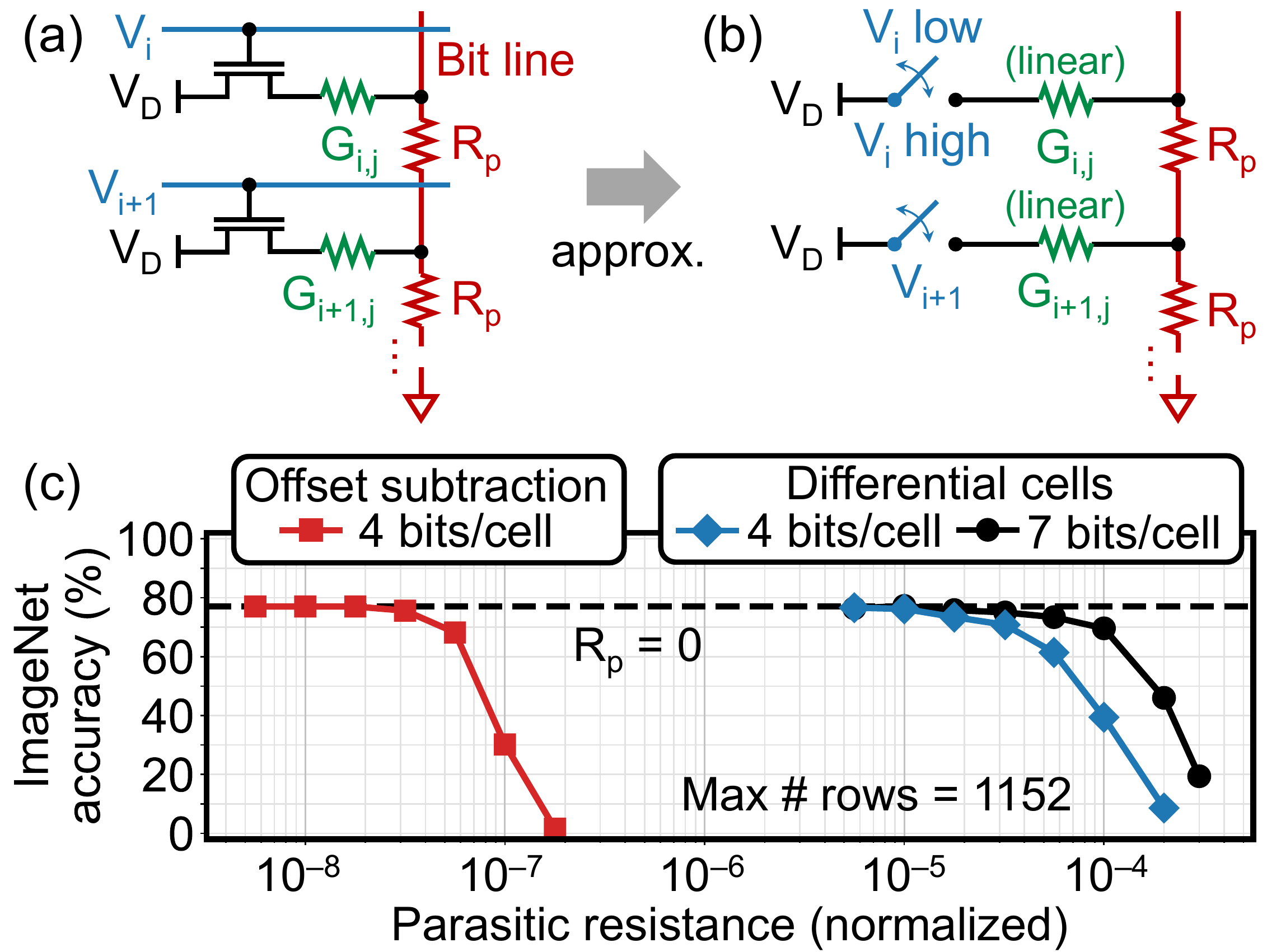}
    \caption{(a) Modeled unit cell and (b) its approximation. (c) ImageNet accuracy using ResNet50-v1.5 (500 images) versus normalized parasitic resistance.}
    \label{fig:parasitics_accuracy}
   \vspace{-8pt}
\end{figure}

To a greater degree than cell errors or ADC errors, errors induced by parasitic resistance depend on the specific array topology. The following analysis assumes the memory cell in Fig.~\ref{fig:parasitics_accuracy}(a), which is the same as that in Fig.~\ref{fig:MVM}(c). Input bits are applied to the gates of select transistors that draw nearly zero current\cite{Bojnordi16,Fick17,Agrawal20,Chen18}. All cells source current from a low-resistance power distribution network ($V_\text{D}$). Thus, only the parasitic resistance of the bit line is considered, whose value between two adjacent cells is denoted $R_p$. The bottom of the bit line is held at virtual ground by the peripheral circuitry, such as a current integrator\cite{Marinella18} or transimpedance amplifier\cite{Li2018}. For computational tractability, neural network simulations use the approximate circuit in Fig.~\ref{fig:parasitics_accuracy}(b). Select transistors are modeled as ideal switches, and a small-signal approximation is made to model the memory devices as linear resistors. To generalize to any memory cell or metal interconnect technology, the sensitivity analysis uses the normalized parasitic resistance $\hat{R}_p$, defined as the ratio of $R_p$ to the minimum cell resistance $1/G_\text{max}$.

When applying inputs one bit at a time, the effect of parasitic resistance varies considerably with bit position. This is because activations are typically skewed heavily toward low values, which makes the higher bits more sparse\cite{Miyashita16}. The lower bits have less sparsity, activate more rows, and have the largest parasitics-induced errors. However, these errors are suppressed to some degree by the input S\&A operation.

Fig.~\ref{fig:parasitics_accuracy}(c) shows the sensitivity of ImageNet accuracy to parasitic resistance for three different weight mapping schemes. The array is limited to at most 1152 rows. The offset subtraction case is more than two orders of magnitude more sensitive to parasitic resistance than differential cells. This large difference can be attributed to three causes. First, due to the lack of proportional mapping, the offset case has a much higher bit line currents (see Fig.~\ref{fig:currents}) and thus larger parasitic voltage drops. Second, for the same voltage drops, cells with high conductance contribute a larger error current to the bit line than cells with low conductance. Third, in the differential case, parasitic resistance perturbs the current on both the positive and negative bit lines in the same direction: downward. When these currents are subtracted, a significant portion of the error induced by the parasitic voltage drops cancels. In the offset case, this cancellation does not occur since the subtracted offset is computed digitally. For the systems that use differential cells, the case with 4 bits/cell is slightly more sensitive to parasitic resistance than unsliced weights. This is because the lower bit slice has higher conductances due to its lack of proportionality (see Fig. \ref{fig:conductance_stats}), making it more sensitive to parasitic resistance.

Using differential cells, the accuracy loss is negligible for $\hat{R}_p \leq 10^{-5}$. This ratio is realistically achieved using cell resistances above $\sim$100 k$\Omega$ and metal interconnects used in scaled process nodes, which can have a resistance of $\sim$1 $\Omega$ per cell in a memory array\cite{Narayanan15}. Analog MVMs can thus be scaled to large arrays ($\sim$1000 rows) without being limited by parasitic resistance.

\section{Case Study: SONOS MVM core}
\label{sec:sonos}

This section demonstrates the design principles outlined in the previous sections using a real memory technology. The case study is a SONOS (silicon-oxide-nitride-oxide-silicon) charge trap memory that has been fabricated in an embedded 40nm process, and for which arrays have been electrically characterized to obtain the cell error properties. This section also examines the effects of the previously described design principles on energy efficiency and area, which can be generalized to other technologies.

\subsection{SONOS Approximate Memory Device}
\label{subsec:sonos}

\begin{figure}[t]
    \centering
    \includegraphics[width=0.5\textwidth]{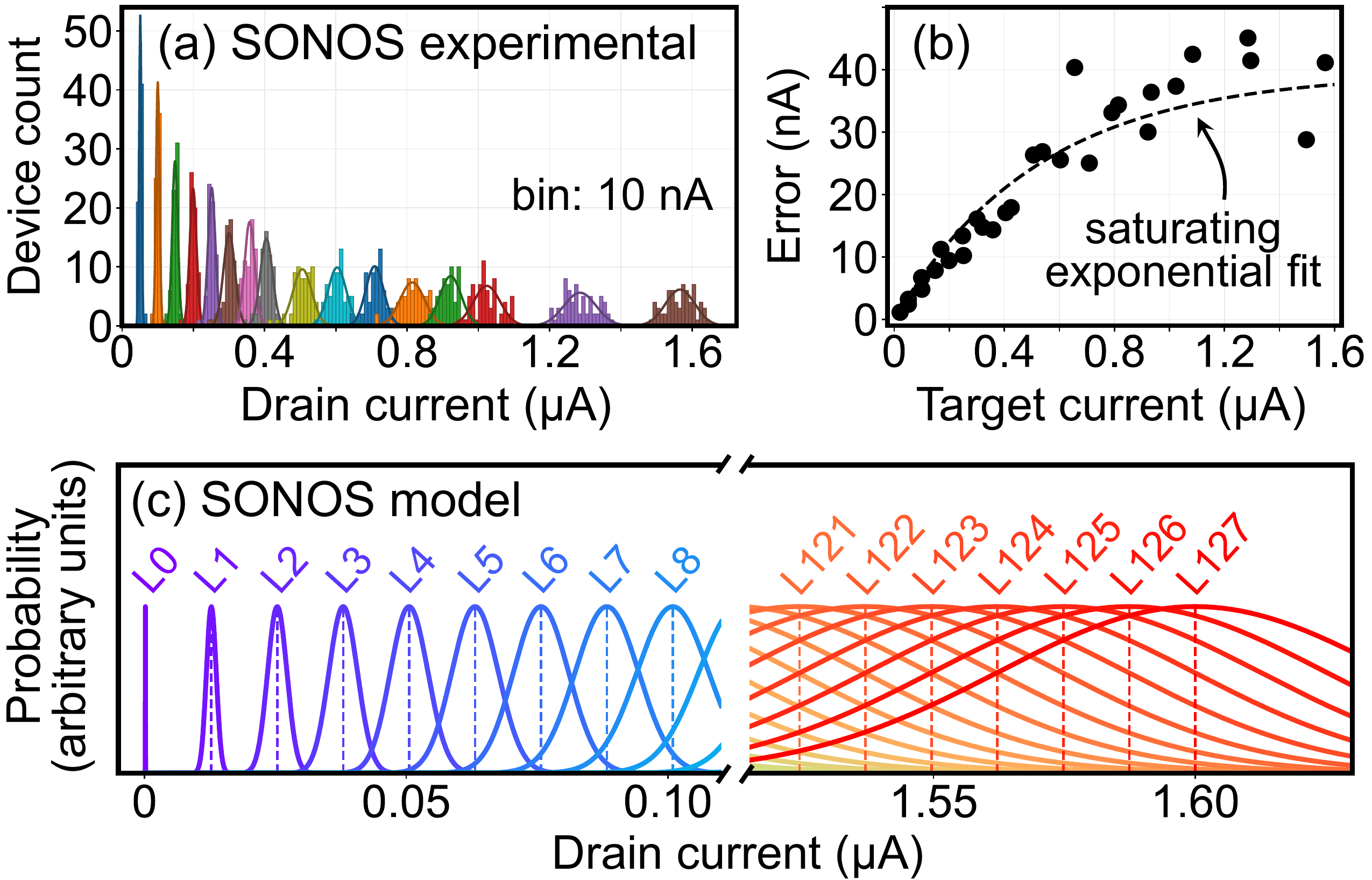}
    \caption{(a) Measured current distributions of SONOS memory cells. Each color is a group of 64 cells programmed to the same target current. (b) Standard deviation of the distribution for various target currents, including histograms not shown in (a). (c) Modeled program error distribution of a 7-bit SONOS cell, based on fit to measurements.
}
    \label{fig:SONOS}
   \vspace{-4pt}
\end{figure}

The two-transistor SONOS flash memory cell has the configuration in Fig. \ref{fig:MVM}(c). The SONOS device is programmed by adding or removing charge from the nitride storage layer, which shifts the threshold voltage $V_\text{T}$ of the transistor channel. The threshold voltage in turn modulates the cell's drain current $I_\text{D}$. The SONOS gate stack and write process were optimized for operation as approximate memory~\cite{Agrawal20,Xiao21}. Histograms of the cell drain current, measured at $V_\text{G} = 0$V and $V_\text{D} = 0.1$V, are shown in Fig.~\ref{fig:SONOS}(a) for various target currents with $I_\text{max} = 1.6$ $\upmu$A. The same biases are used during an MVM. Each histogram is fit to a normal distribution whose width is the expected programming error in a cell.

Fig.~\ref{fig:SONOS}(b) shows the error as a function of current. The cell error is approximately state-proportional below 0.5 $\upmu$A with $\alpha_\text{prop} \approx 6\%$, and saturates at high conductance. This property comes from the fact that the SONOS transistor is designed to operate in the subthreshold regime at a fixed bias of $V_\text{G} = 0$V:
\begin{align}
I_\text{D} = I_0 \exp \left( -\eta \frac{qV_\text{T}}{kT} \right)
\end{align}
where $I_0$ is a constant, $\eta$ is the gate efficiency, $q$ is the electron charge, $k$ is the Boltzmann constant, and $T$ is the temperature. Differentiating the above with respect to $V_\text{T}$ gives:
\begin{align}
\label{eq:SONOS}
dI_\text{D} = -\frac{q\eta}{kT} dV_\text{T} \times I_\text{D}
\end{align}
Since the amount of stored charge is related linearly to $V_\text{T}$, the error in the charge injection or removal process is proportional to the error $dV_\text{T}$. Equation \eqref{eq:SONOS} shows that for the same error in the write process, the error in the cell current $dI_\text{D}$ is proportional to the cell current $I_\text{D}$. This is consistent with the data in Fig. \ref{fig:SONOS}(b). At currents above around 0.8 $\upmu$A, corresponding to lower $V_\text{T}$, the device leaves the subthreshold regime and the error consequently increases sublinearly with the current. As discussed in Section \ref{sec:errors}, state-proportional error is highly advantageous for neural network inference, as it matches the most frequently used weight values to devices with the least error.

The state-dependent error of the SONOS device is modeled within CrossSim using a saturating exponential fit to the data, shown in Fig. \ref{fig:SONOS}(b). The SONOS cell is used as an approximate 7-bit memory; the modeled program error distributions of the 128 target current levels are shown in Fig. \ref{fig:SONOS}(c). The device is programmed into deep subthreshold (highest possible $V_\text{T}$) for the lowest state to realize an On/Off current ratio of $10^7$. The ImageNet accuracy with this device will be evaluated in Section \ref{subsec:sonos_accuracy_eval}.

\subsection{MVM Core Design}
\label{subsec:core_design}

\begin{figure}[t]
    \centering
    \includegraphics[width=0.485\textwidth]{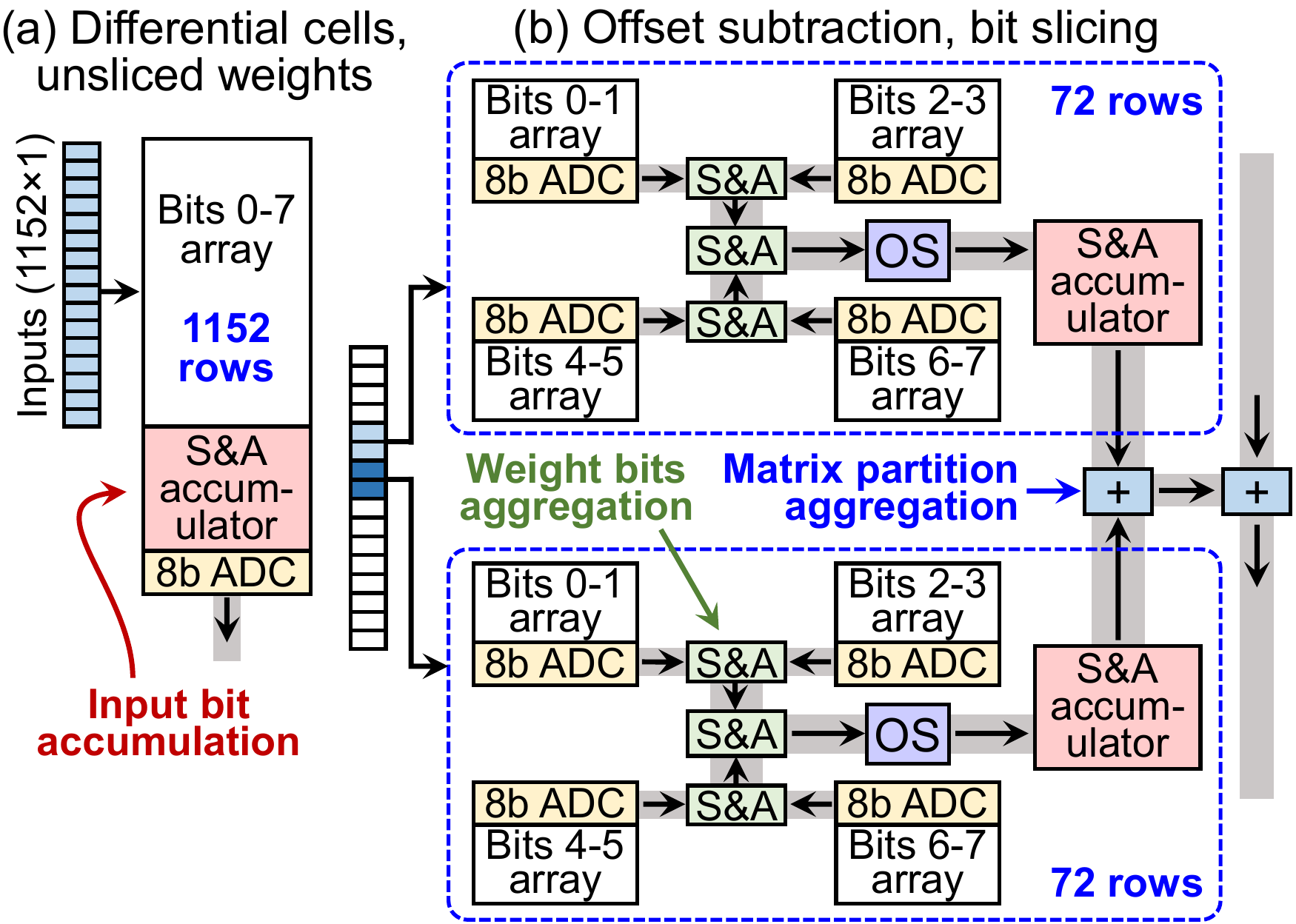}
    \caption{Two core configurations used in the energy and area evaluation. Based on results in Section \ref{sec:ADC}, a matrix with 1152 rows must be partitioned across multiple arrays to maintain accuracy with offset subtraction (OS).
}
    \label{fig:aggregation}
   \vspace{-8pt}
\end{figure}

Since this work addresses the design of the analog core, the energy and area results here will be restricted to the core level for generalizability. A core is defined as the collection of processing elements that perform a full-precision MVM; all bit slices, input bits, and matrix partitions. Fig.~\ref{fig:aggregation} shows the evaluated core design for two example weight mapping schemes. An 8-bit ADC is used for all of the considered design points.

As shown in Section \ref{sec:ADC}, cores that use differential cells can achieve high accuracy with an 8-bit ADC independent of the array size and bits per cell. Therefore, both parameters can be swept without affecting accuracy. A maximum array size of 1152 rows is assumed to control the parasitic voltage drops, based on the results in Section \ref{sec:parasitics} for ResNet50-v1.5.

Offset-subtraction cores have a more limited design space. Fig. \ref{fig:size_accuracy} shows that due to 8-bit ADC quantization alone, offset subtraction can reach high accuracy only with a small array ($\leq$144 rows) and finer bit slices ($\leq$2 bits/cell). Sections \ref{sec:errors} and \ref{sec:parasitics} showed that they are also more sensitive to cell errors and parasitic resistance, which might further reduce the array size. Additionally, the absence of proportional mapping requires larger and more power-hungry peripheral circuits that can support larger bit line currents, as shown in Fig. \ref{fig:currents}. For these reasons, only four design points are evaluated for offset subtraction, both using digital input bit accumulation: 72 and 144 rows with 1 and 2 bits per cell, which are close to the design point in ISAAC\cite{Shafiee16}. One of these designs is shown in Fig.~\ref{fig:aggregation}(b) and requires multiple digital steps to aggregate partial results produced by the analog hardware.

All energy and area estimates are based on SONOS arrays and peripheral circuits that are designed and simulated in an embedded 40nm process compatible with SONOS memory\cite{Kouznetsov2018,Agrawal20}. The energy consumption of the array and row drivers is based on the average cell conductances in Fig. \ref{fig:conductance_stats} and the average activity factors for each input bit when running ResNet50-v1.5 on ImageNet. The core uses a current conveyor that integrates each input bit for 10 ns\cite{Marinella18}, a switched-capacitor circuit for analog input bit accumulation~\cite{Bavandpour20}, and a power- and area-efficient 8-bit ramp ADC clocked at 1 GHz~\cite{Marinella18}. ADC range calibration is implemented with a tunable operational-amplifier gain stage after the integrator. Digital component energies are derived from a standard cell library. Since all array outputs are simultaneously available with a ramp ADC, as many S\&A units are allocated as needed to process these results in parallel. Area is estimated from the sum of circuit block areas rather than a physical layout.

\subsection{Energy and Area Evaluation}
\label{subsec:sonos_energy_eval}

\begin{figure}[t]
    \centering
    \includegraphics[width=0.485\textwidth]{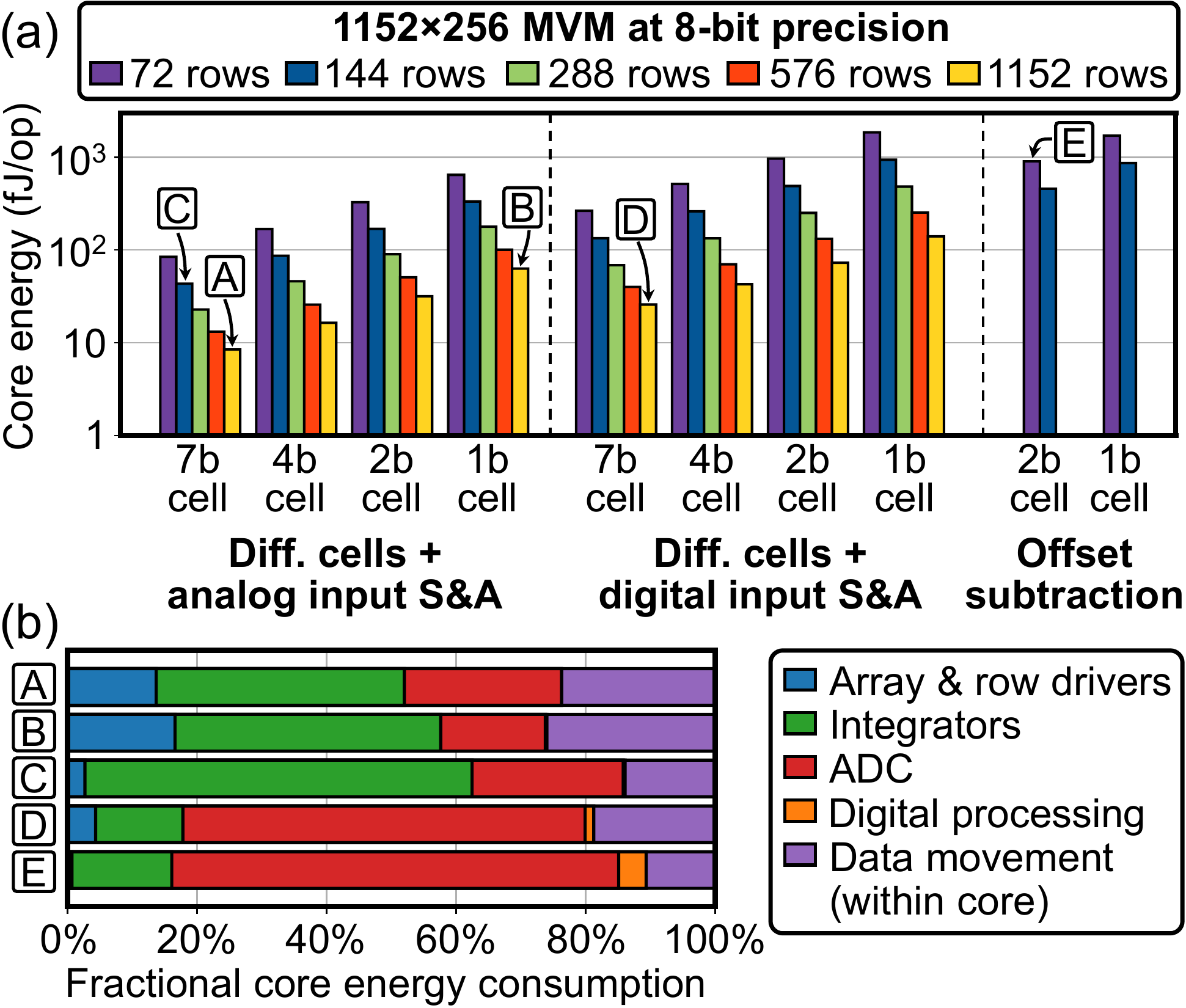}
    \caption{(a) Core energy per operation (1 MAC = 2 operations) for various core configurations applied to an MVM of size 1152 $\times$ 256. (b) Breakdown of energy use among core components for selected configurations.
}
    \label{fig:energy}
   \vspace{-8pt}
\end{figure}

Fig.~\ref{fig:energy}(a) shows the energy efficiency of various core configurations. To estimate the peak efficiency, a $1152 \times 256$ weight matrix is evaluated that utilizes every cell in each array. Table \ref{tab:energy} details the area and energy efficiency of five labeled configurations in Fig.~\ref{fig:energy}(a), whose energy breakdown among core components is shown in Fig.~\ref{fig:energy}(b). These results reveal several trends: 

(1) Unsliced weights are more efficient than bit slicing because the bit line peripheral circuit costs increase roughly linearly with the number of slices. The area increases linearly with bit slicing due to having more cells per weight.

(2) Larger arrays are more efficient since the integrator and ADC energies are amortized over more operations, and less computation is done in the less efficient digital domain. Density also improves since these circuits are shared by more matrix elements. While smaller arrays can offer better area utilization when mapping small matrices \cite{Nag18,Sze19}, large arrays are necessary to extract the efficiency benefits of analog processing. Neural networks that more fully utilize large arrays will have superior system-level energy efficiency when deployed in an analog accelerator.

(3) Analog input bit accumulation yields a $2$-$4\times$ energy improvement. The technique increases integrator energy, but reduces the number of ADC conversions by $8\times$. When each input bit requires a digitization step, the ADC dominates the energy cost, as shown in Fig.~\ref{fig:energy}(b) for designs D and E; this is consistent with prior work~\cite{Shafiee16,Chi16}.

For the reasons summarized in Section \ref{subsec:core_design}, systems that rely on offset subtraction cannot exploit any of the above techniques to reduce energy. Design A is the most efficient design for differential cells, while Design E is an offset-subtraction design that very nearly satisfies the FPG (a pre-requisite for high accuracy using offset subtraction, as explained in Section \ref{subsec:ADC_precision}). Design E has 107$\times$ higher energy consumption and $46\times$ larger area. The higher energy comes from having $4\times$ as many bit slices, $8\times$ as many ADC conversions per input value, and $16\times$ as many arrays to map a large matrix. The actual ratio of energy consumption is smaller than the product of these factors since only part of the total energy scales with these factors.

\renewcommand{\arraystretch}{1.15}
\begin{table}[t]
    \caption{Efficiency of selected core configurations}
    \label{tab:energy}
    \setlength\tabcolsep{0pt}
    \setlength\doublerulesep{5pt}
    \begin{threeparttable}
    \begin{tabularx}{0.49\textwidth}{|>{\hsize=2.2\hsize}X|>{\hsize=0.76\hsize}X|>{\hsize=0.76\hsize}X|>{\hsize=0.76\hsize}X|>{\hsize=0.76\hsize}X|>{\hsize=0.76\hsize}X|}
        \hline
        \centering \textbf{Design} & \centering\arraybackslash\textbf{A}& \centering\arraybackslash \textbf{B} & \centering\arraybackslash \textbf{C} & \centering\arraybackslash \textbf{D} & \centering\arraybackslash \textbf{E} \\
        \hline
        \centering \textbf{Negative values} & \centering Diff. & \centering\arraybackslash Diff. & \centering\arraybackslash Diff. & \centering\arraybackslash Diff. & \centering\arraybackslash Offset \\
        \hline
        \centering \textbf{Weight resolution} & \centering 8 & \centering\arraybackslash 9 & \centering\arraybackslash 8 & \centering\arraybackslash 8 & \centering\arraybackslash 8 \\
        \hline
        \centering \textbf{Bits / cell} & \centering 7 & \centering\arraybackslash 1 & \centering\arraybackslash 7 & \centering\arraybackslash 7 & \centering\arraybackslash 2 \\
        \hline
        \centering \textbf{\# rows} & \centering 1152 & \centering\arraybackslash 1152 & \centering\arraybackslash 144 & \centering\arraybackslash 1152 & \centering\arraybackslash 72 \\
        \hline
        \centering \textbf{Input bit S\&A}\tnote{$\dagger$} & \centering A & \centering\arraybackslash A & \centering\arraybackslash A & \centering\arraybackslash D & \centering\arraybackslash D \\
        \hline
        \centering \textbf{ADC resolution} & \centering 8 & \centering\arraybackslash 8 & \centering\arraybackslash 8 & \centering\arraybackslash 8 & \centering\arraybackslash 8 \\
        \hline
        \centering \textbf{Ideal analog resolution} $B_\text{out}$ & \centering 26.2 & \centering\arraybackslash 20.2 & \centering\arraybackslash 23.2 & \centering\arraybackslash 18.2 & \centering\arraybackslash 8.2 \\
        \hline
        \hline
        \centering \textbf{Core area} (mm$^2$) & \centering 0.24 & \centering\arraybackslash 2.02 & \centering\arraybackslash 1.30 & \centering\arraybackslash 0.27 & \centering\arraybackslash 11.14 \\
        \hline
        \centering \textbf{Core energy} (fJ/op) & \centering 8.4 & \centering\arraybackslash 63.1 & \centering\arraybackslash 43.3 & \centering\arraybackslash 25.8 & \centering\arraybackslash 902.0 \\
        \hline
    \end{tabularx}

    \vspace{2pt}
    \footnotesize
    $^\dagger$A = analog, D = digital.
    \end{threeparttable}

    \caption[justification=centering]{ResNet50-v1.5 accuracy with SONOS errors using selected core designs (1000 images, 10 runs each)}
    \label{tab:sonos_accuracy}
    \begin{tabularx}{0.49\textwidth}{|>{\hsize=1.4\hsize}X|>{\hsize=0.92\hsize}X|>{\hsize=0.92\hsize}X|>{\hsize=0.92\hsize}X|>{\hsize=0.92\hsize}X|>{\hsize=0.92\hsize}X|}
        \hline
        \centering \textbf{Design} & \centering\arraybackslash\textbf{A}& \centering\arraybackslash \textbf{B} & \centering\arraybackslash \textbf{C} & \centering\arraybackslash \textbf{D} & \centering\arraybackslash \textbf{E} \\
        \hline
        \centering Ideal cells & \centering 76.3\% & \centering\arraybackslash 75.3\%  & \centering\arraybackslash 75.4\% & \centering\arraybackslash 76.3\% & \centering\arraybackslash 74.9\% \\
        \hline
        \centering SONOS & \centering 74.0\% \linebreak $\pm$1.0\% & \centering\arraybackslash 75.4\% \linebreak $\pm$0.3\%  & \centering\arraybackslash 73.6\% \linebreak $\pm$0.7\% & \centering\arraybackslash 74.1\% \linebreak $\pm$1.0\%  & \centering\arraybackslash  50.2\% $\pm$5.3\% \\
        \hline
    \end{tabularx}
    \vspace{-8pt}

    \caption[justification=centering]{Accuracy of the analog inference accelerator on the full ImageNet test set (50,000 images, 10 runs each)}
    \label{tab:sonos_accuracy_2}
    \begin{tabularx}{0.49\textwidth}{|>{\hsize=1.2\hsize}X|>{\hsize=0.9\hsize}X|>{\hsize=0.9\hsize}X|}
        \hline
        \centering\arraybackslash\textbf{ResNet50-v1.5} & \centering\arraybackslash\textbf{Floating-point}& \centering\arraybackslash \textbf{4-bit, QAT}  \\
        \hline
        \centering\arraybackslash Fully digital & \centering\arraybackslash 76.466\% & \centering\arraybackslash 76.154\%  \\
        \hline
        \centering\arraybackslash Design A, ideal cells & \centering\arraybackslash 76.082\% & \centering\arraybackslash 76.038\%  \\
        \hline
        \centering\arraybackslash Design A, SONOS & \centering\arraybackslash 74.296\% \linebreak$\pm$0.348\% & \centering\arraybackslash 75.294\% \linebreak$\pm$0.192\%  \\
        \hline        
    \end{tabularx}
    \vspace{-8pt}
\end{table}

\subsection{Accuracy Evaluation}
\label{subsec:sonos_accuracy_eval}

Table \ref{tab:sonos_accuracy} compares the ImageNet accuracy with ResNet50-v1.5 obtained using the same five design points. The simulations include 8-bit weight and activation quantization, 8-bit ADCs calibrated separately for each design, and random SONOS programming errors following the full state dependence in Fig. \ref{fig:SONOS}(b), sampled ten times as described in Section \ref{subsec:devices}. The small differences in the baseline accuracy using ideal cells result from the varying effectiveness of the 8-bit ADC calibration across designs.

To keep the computations tractable, parasitic resistance was not included. Relative to the SONOS cells, the metal interconnects in the 40nm process have a normalized resistance of $\hat{R}_p \approx 10^{-5}$. Fig. \ref{fig:parasitics_accuracy} shows that this resistance has negligible effect on the accuracy of Designs A, C, and D, which use differential cells and unsliced weights. For the other designs, the accuracies in Table \ref{tab:sonos_accuracy} are best-case estimates; with a realistic parasitic resistance, the accuracy of Design B may be slightly lower, and that of Design E is likely to be much lower.

The designs with differential cells and unsliced weights (A, C, and D) all have similar accuracies, losing roughly 2\% on ImageNet by using SONOS cells. Design B, which uses 1-bit slices, is less sensitive to SONOS errors than unsliced weights. This result is consistent with Fig. \ref{fig:proportional_error}(b), and is due to the fact that finer bit slicing creates greater sparsity in the most significant slice, as explained in Section \ref{subsec:proportional_error}. However, this design requires nearly $8\times$ larger energy and area than Design A. Whether this small difference in accuracy is worth the considerable overhead is dependent on the end-to-end application requirements.

Design E loses more than 20\% in accuracy from SONOS cell errors. This design is the least robust because it uses offset subtraction, which lacks weight proportionality and thus does not exploit the state-proportional error property of the cells. At the average cell current used by this system ($0.5I_\text{max}$, or 0.8 $\upmu$A), the SONOS error properties are intermediate between state-independent error and state-proportional error (with $\alpha_\text{ind} \approx \alpha_\text{prop} \approx 4\%$). Notably, however, the accuracy of this design is much higher than that predicted in Fig. \ref{fig:independent_error}(a) or Fig. \ref{fig:proportional_error}(a) for offset subtraction with 2 bits/cell. This is due to the ADCs, which cut off the analog accumulation of cell errors beyond 72 rows: this effect is depicted in Fig. \ref{fig:coarse_quant}(b). As noted earlier, the true accuracy of Design E is likely much lower than listed in Table \ref{tab:sonos_accuracy} due to parasitic resistance.

Table \ref{tab:sonos_accuracy_2} shows the accuracy of the most efficient design, Design A, on the full ImageNet test set of 50,000 images. The 2.17\% accuracy loss on ResNet50-v1.5 is relatively small for a system that uses direct weight transfer. By comparison, the PCM devices in Joshi \etal{} lose 7.8\% ImageNet accuracy using ResNet34\cite{Joshi20}. The main accuracy advantage of the SONOS device over PCM is state-proportional error, as explained in Section \ref{subsec:sonos}. Table \ref{tab:sonos_accuracy_2} further shows that by using a standard quantization-aware training scheme with 4-bit activations, the propagation of this error can be significantly suppressed as described in Section \ref{sec:quant}. This reduces the accuracy loss induced by the SONOS device to only 0.86\%.

\section{Conclusions}
\label{sec:conclusion}

Error resilience can be built into analog accelerators by designing the system to leverage the properties of the application neural network. A proportional mapping of numerical values in the algorithm to physical quantities in the accelerator exploits a feature common to many networks: a weight distribution that is skewed toward low values. This paper showed that a proportional mapping reduces sensitivity to several categories of analog errors. The critical building blocks of a proportional system are differential cells for mapping signed weights, a memory technology with high On/Off ratio, and programming errors that scale with conductance.

This paper also evaluated the popular design choices made by prior analog accelerators from the perspective of accuracy and robustness to errors. Bit slicing has only a small accuracy benefit, which is unlikely to outweigh the considerable energy and area overhead needed to support it. The full-precision guarantee is also too conservative a choice for neural network inference, and leads to smaller arrays or greater ADC overheads than needed. Proportional systems can perform a much larger share of the computation in analog, and allow the algorithm to dictate the precision with which the analog outputs are digitized. 

In analog systems, where algorithmic accuracy depends on device-level effects, hardware design should ultimately be guided by a rigorous evaluation of the \emph{end-to-end} accuracy. While the evaluation of a design choice on the basis of intermediate results (such as MVM-level precision) can yield valuable insights, an end-to-end accuracy evaluation is needed to avoid unnecessary bottlenecks for accuracy and efficiency. An end-to-end design approach results in a harmonization of the hardware and the algorithm that ultimately delivers the order-of-magnitude energy efficiency benefits promised by analog accelerators.

\section*{Acknowledgments}
\label{sec:acknowledgments}

This work was supported by the Laboratory Directed Research and Development program at Sandia National Laboratories, a multimission laboratory managed and operated by National Technology and Engineering Solutions of Sandia LLC, a wholly owned subsidiary of Honeywell International Inc. for the U.S. Department of Energy’s National Nuclear Security Administration under contract DE-NA0003525. This paper describes objective technical results and analysis. Any subjective views or opinions that might be expressed in the paper do not necessarily represent the views of the U.S. Department of Energy or the United States Government.


\end{document}